\documentclass[superscriptaddress,secnumarabic,amssymb,amsmath,nobibnotes,aps,prd,showpacs,nofootinbib]{revtex4}
\usepackage{caption}
\usepackage{subcaption}
\usepackage{graphicx}
\usepackage{bm}
\usepackage{amsmath}
\usepackage{amsfonts}
\usepackage{amssymb}
\usepackage{epstopdf}
\usepackage{color}%
\providecommand{\U}[1]{\protect\rule{.1in}{.1in}}
\newcommand{\be}{\begin{equation}}
\newcommand{\ee}{\end{equation}}

\newcommand{\mincir}{\raise
-3.truept\hbox{\rlap{\hbox{$\sim$}}\raise4.truept\hbox{$<$}\ }}
\newcommand{\magcir}{\raise
-3.truept\hbox{\rlap{\hbox{$\sim$}}\raise4.truept\hbox{$>$}\ }}

\newtheorem{remark}{Remark}[section]

\begin{document}

\title{Bouncing cosmologies via modified gravity in the ADM formalism: Application to Loop Quantum Cosmology}

\author{Jaume de Haro\footnote{E-mail: jaime.haro@upc.edu}}
\affiliation{Departament de Matem\`atiques, Universitat Polit\`ecnica de Catalunya, Colom 11, 08222 Terrassa, Spain}
\affiliation{Departament de Matem\`atiques, Universitat Polit\`ecnica de Catalunya, Diagonal 647, 08028 Barcelona, Spain}

\author{Jaume Amor\'os\footnote{E-mail: jaume.amoros@upc.edu}}
\affiliation{Departament de Matem\`atiques, Universitat Polit\`ecnica de Catalunya, Diagonal 647, 08028 Barcelona, Spain}



\thispagestyle{empty}

\begin{abstract}

We consider the ADM formalism as a tool to build bouncing cosmologies. In this approach, the foliation of the spacetime has to be fixed  in order to go beyond General Relativity modifying the gravitational
sector. Once a preferred slicing,  which we choose based on the matter content of the universe following the spirit of Weyl's postulate, has been fixed,
$f$ theories depending on the extrinsic and intrinsic curvature of the slicing are covariant for all the reference frames preserving the foliation, i.e., the constraint and dynamical equations have the 
same form 
for all these observers. 
Moreover, choosing multivalued $f$ functions, bouncing backgrounds emerge in a 
natural way. In fact, the simplest is the one corresponding to holonomy corrected Loop Quantum Cosmology. The final goal of this work is to provide the equations of perturbations which,
unlike the full equations,  become gauge invariant in this theory,  and apply
them to the so-called matter bounce scenario.

\end{abstract}

\vspace{0.5cm}

\pacs{04.20.Fy, 04.50.Kd, 98.80.Jk.}

\maketitle

\hspace{1.4 cm} {{ Keywords:}  Modified gravity, ADM formalism, Loop Quantum Cosmology, Weyl postulate.}


\section{ Introduction}
Bouncing cosmologies (see the reviews \cite{nb08,b12,b09,b10,cai})  seem to be the most natural alternative to the inflationary  paradigm \cite{guth, linde,steinhardt}, 
although  in some papers \cite{bct, bojo,wang,wang1} it seems to be only an implementation of it, introducing a pre-inflationary dynamics in order  to remove the initial singularity.  However,  it is not an easy task to build
up this kind of scenarios
in the framework  of the General Relativity (GR)  \cite{allen, cqplz, qeclz, lbp, rs},
or to go beyond GR  using modified theories such as  $f(R)$ gravity   (see for example \cite{bmmno, noo,oo}), modified Gauss-Bonnet gravity \cite{bmmo, bmmo1, noo1}  
or Born-Infeld inspired gravity theories \cite{beltran,beltran1}.
A simple non-singular bounce, which replaces the Big Bang singularity, emerges
from holonomy corrected Loop Quantum Cosmology (LQC)  \cite{aps06a, singh06,svv06,singh08,cm10,dmp09,singh09, sst06, nw07,cs09}.

\

{
The LQC theory,   in the flat Friedmann-Lema{\^\i}tre-Robertson-Walker (FLRW)
 spacetime,  containing only holonomy corrections
(not inverse-volume effects) can be built as follows: 
since the gravitational part of the classical Hamiltonian contains the Ashtekar connection,  namely ${\mathfrak c}=\beta \dot{a}$, where $\beta$ is the Immirzi parameter  \cite{immirzi} and $a$ the scale factor, 
 which does not have a well defined quantum operator (see for instance \cite{as,bho}), so
one has to express 
it in terms of holonomies in order to obtain a quantity able to be quantized which captures the underlying loop quantum dynamics (see for instance \cite{aps06}). At effective level, as has been shown in \cite{dmp, he}, this is equivalent
to replacing the Ashtekar connection 
by the function $\sin({\mu}{\mathfrak c})/{\mu}$, where $\mu$ is a parameter with dimensions of length. Then, with the new
holonomy corrected Hamiltonian constraint one obtains the modified LQC-Friedmann equation, which  in the phase space $(H,\rho)$ represents an ellipse \cite{aho}, where $H$ is the Hubble parameter and 
$\rho$ the energy density, and thus, contrarily to GR where this curve is an unbounded parabola,  singularities such as the big bang,  big rip  \cite{k} or little rip \cite{frampton}  are 
directly removed.

}

\

Given the success of the effective
theory for homogeneous and isotropic space-times, it is natural to try
the same approach for cosmological perturbations in LQC. If one simply
replaces the Ashtekar connection everywhere it appears in the
Hamiltonian constraint by a sinusoidal function    $ {\mathfrak c}\rightarrow \frac{\sin(n {\mu}{\mathfrak c})}{n{\mu}} $  where $n$ is a natural number \cite{bh}, the result is a theory with
anomalies in its constraint algebra. A more careful treatment that
gives an anomaly-free theory in the longitudinal gauge has been found
\cite{wilson-ewing}, and this has since been extended to the gauge-invariant case by
allowing for new terms in the holonomy modified Hamiltonian that
vanish in the classical limit \cite{cmbg, gbg,cbvg,clb}). The result is a modified
Mukhanov-Sasaki equation that can be used to study the dynamics of
cosmological perturbations in LQC.

\

\

On the other hand, it is well-known that the background of holonomy corrected LQC in flat FLRW spacetime can be easily mimicked in modified gravity if one has an scalar field that only depends on the
square of the Hubble parameter \cite{helling,ds09,haro12}. This scalar appears in teleparallelism, where the spacetime is equipped by the Weitzenb\"ock connection \cite{w}
and an orthonormal basis parallelaly transported is chosen in the tangent bundle 
(see for example \cite{hvkn,ff08}). For a FLRW geometry, using  synchronous co-moving coordinates,
 the scalar
named torsion and denoted by ${\mathcal T}$, is given by ${\mathcal T}=-6H^2$ \cite{bf08,ha}. Then, the $f({\mathcal T})$ theory that leads to the same ellipse as holonomy corrected LQC could be 
named teleparallell LQC \cite{haro13}. Dealing with 
perturbations, the perturbed equations in teleparallelism \cite{cdds,ccdds} have been applied to this particular function in \cite{ha14,ha14a} obtaining a different kind of equation, which differs 
from the ones of
holonomy corrected LQC starting with the velocity of the speed of sound. Unfortunately, besides avoiding the usual Levi-Civita connection, 
the scalar torsion is not local Lorentz invariant, that is, 
a local Lorentz transformation of  the orthonormal basis could lead to a new orthogonal basis, in which the value of the scalar torsion is different form the one obtained using the first basis  \cite{lsb, slb}.
Thus,  given a torsion, a $f({\mathcal T})$ theory  becomes  preferred frame,
which is an undesirable feature. Effectively, let $\{{\bf e}_{\nu}\}_{\nu=0,..,3}$ and $\{\bar{\bf e}_{\nu}\}_{\nu=0,..,3}$ be two orthonormal basis in the tangent bundle parallelally transported, and let 
${\mathcal T}$ and $\bar{\mathcal T}$ be their corresponding torsions. Since its difference is a divergence,  when $f$ is the identity their corresponding actions are the same, which will not happen
for a general function  $f$.

\

In the same way as in teleleparallelism, in the Arnowitt-Deser-Misner (ADM) formalism of GR \cite{ADM, Gourgoulhon}, where a globally hyperbolic spacetime could be foliated by three-dimensional space-like sub-manifolds 
parameterized by a {\it time} and 
the connection is the usual Levi-Civita one,
the extrinsic curvature scalar, namely 
${\mathcal I}$, is also equal to $-6H^2$ in the flat FLRW spacetime using  synchronous  co-moving coordinates. Since the scalar curvature 
of the full manifold, namely $R$, is equal to ${\mathcal R}+{\mathcal I}$ plus a divergence, where ${\mathcal R}$ is the intrinsic curvature, i.e., the scalar curvature of the sub-manifolds,
 on can built some classes of $F({\mathcal R}, {\mathcal I})$ gravitational theories, leading to different backgrounds and one of them will be the
same as in holonomy corrected  LQC. In this way, the main goal of this work is to provide the gravitational equations in 
a particular case, namely $F({\mathcal R}, {\mathcal I})= {\mathcal R}+ f({\mathcal I})$, its perturbed equations and, as a particular case, to calculate explicitly them for the $f$ 
theory that
leads to the same background as LQC.
At this point, it is important to emphasize that the theory is gauge dependent, in the sense that, it depends on the foliation chosen or, equivalently, on the choice of the
normal
vector, because only the sum of the intrinsic
and extrinsic
curvature is the same, up a divergence, for all the slicing. So, when the gauge is fixed, the theory becomes covariant only for the coordinate systems preserving the foliation, i.e.,  
the equations only will become with the same aspect  in these reference frames. Since cosmology
deals with the evolution of the universe, in order to have a well defined Cauchy problem \cite{Gourgoulhon,smarr}   one has to fix precisely a foliation,  and of course there are some gauges such as
 the so-called ``geodesic slicing''  \cite{bardeen},  
where the lapse function is chosen to be $1$,
   the constant main curvature (the trace of the extrinsic curvature tensor) slicing \cite{marsden},  the   maximal slicing  \cite{ murchadha},
   which corresponds to the vanishing of the mean curvature of the hyper-surfaces  $\Sigma_t$,    or the harmonic slicing  \cite{donder}, where the $t$ coordinate is required to be harmonic
   with respect to the  D'Alembertian. In this work, as in  the Chapter  $4$ of \cite{ellis}, and following the spirit of Weyl's principle
   (see \cite{rz} for a historical and critical review), we argue that there is  a preferred 
   $4$-velocity field in the whole spacetime, which generates a preferred non-crossing family of world-lines. Therefore, taking
   the parameter $t$ as the proper time 
of these preferred observers one obtains a preferred {\it synchronous gauge},   or choosing 
the time $t$
 defining the so-called {\it co-moving slicing}, i.e., the slicing  orthogonal to the world-lines \cite{riotto}, which, as we will see, is the 
best choice when the universe is filled by an scalar field, 
one obtains two preferred different $3+1$ splittings of the spacetime.
Fortunately, and it is a nice feature of the   theory, the equations of perturbations are gauge invariant, i.e.,  slicing independent, provided that the background, which  is the FLRW spacetime, 
has been split 
as in GR, using co-moving coordinates.

\ 

The manuscript is organized as follows: In section II we review the ADM formalism and, for a given foliation, we generalize GR to a ${\mathcal R}+f({\mathcal I})$ gravitational theory, obtaining the particular $f$ that leads to the same background as  holonomy corrected LQC.
At the end of the Section, the  gravitational equations that generalize those of GR are derived. Section III is devoted to the study of scalar perturbations in this theory, 
comparing them with those of the teleparallel $f({\mathcal T})$ gravity, 
and also  applying them to LQC. In section IV we do the same as in
section III but for tensor perturbations,  and finally, in section V  we apply our previously obtained results to the matter bounce scenario in LQC.

\

The units used throughout the paper are $\hbar=c=1$, where $M_{pl}=\frac{1}{\sqrt{8\pi G}}$ is the reduced Planck's mass.

\section{ADM formalism revisited}

In order to deal with bouncing cosmologies without initial or final singularities we assume that the spacetime is 
globally hyperbolic, i.e., it admits  a three-dimensional space-like Cauchy surface $\Sigma$, four-dimensional Lorentz manifold ${\mathcal M}$ with metric $g$ (see \cite{haw} or the Chapter $3$ of 
\cite{Gourgoulhon} for a review of these
concepts), and thus,
following the Geroch's splitting theorem \cite{geroch},
topologically the spacetime is  homeomorphic to $\Sigma\times \mathbb{R}$ and thus it admits a {\it foliation or slicing} 
 $(\Sigma_t)_{t\in \mathbb{R}}$ where for $t\in \mathbb{R}$
 \begin{eqnarray}
  \Sigma_t\equiv \{p\in {\mathcal M}: \hat{t}(p)=t\},
 \end{eqnarray}
being $\hat{t}:{\mathcal M}\rightarrow \mathbb{R}$ a scalar field defined in the full spacetime.

 \
 
 As in the  Arnowitt-Deser-Misner  (ADM)  formulation of GR \cite{ADM},  we make the decomposition $\partial_t=N{\bf n}+ {\bf N}$ where $N$ 
is the so-called lapse function, ${\bf N}$ is a vector belonging in the tangent space of $\Sigma_t$  named the shift vector and ${\bf n}$ is a vector perpendicular to $\Sigma_t$, satisfying
$g({\bf n,\bf n})=-1$. Now,  let $\{\partial_j\}_{j=1,2,3}$  be a basis of  the tangent space of $\Sigma_t$,
 then the metric, in coordinates, is given by
 \begin{eqnarray}
 ds^2=g_{\mu\nu}dx^{\mu}dx^{\nu}= -N^2dt^2+\gamma_{ij}(dx^i+N^idt)(dx^j+N^jdt),
 \end{eqnarray}
 where $\gamma_{ij}=g(\partial_i,\partial_j)$ are the entries of the induced metric in $\Sigma_t$, which we denote by $\gamma$, and $N^i$ are the coordinates
 of the vector ${\bf N}$ in the basis $\{\partial_j\}_{j=1,2,3}$, i.e., 
 ${\bf N}=N^j\partial_j$.
 
 \
 
 An important ingredient of this formulation is the extrinsic curvature tensor  which, although defined with a negative sign in some books, in this work is given by
 \begin{eqnarray} 
 K({\bf u},{\bf v})=
 g(\nabla_{\bf u} {\bf v}, {\bf n}),
 \end{eqnarray}
 where we have denoted by  $\nabla$  the  Levi-Civita connection, and
 ${\bf u}, {\bf v}$ are vectors of the tangent space of  $\Sigma_t$.   
 
  \

 Let  ${\mathcal R}$ be the intrinsic curvature, i.e., the scalar curvature of $\Sigma_t$ , and ${\mathcal I}= K_{ij}K^{ij}-(Tr(K))^2$, where $Tr(K)=K_i^i$ is the trace of the extrinsic curvature tensor. Then, 
 the scalar curvature of the 
 whole spacetime ${\mathcal M}$  is related with these quantities via the equality \cite{Gourgoulhon}
 \begin{eqnarray}
R= {\mathcal R}+{\mathcal I}-2 \nabla_i v^i,
\end{eqnarray} 
 where $ \nabla_i v^i, $ is the divergence of the vector field ${\bf v}=\nabla_{\bf n} {\bf n}-\nabla_i n^i{\bf n}$.   
 
 Therefore, the gravitational part of the Hilbert-Einstein action could be written as 
 \begin{eqnarray}
 S_{HE; grav}=\frac{M_{pl}^2}{2}\int_{{\mathcal M}} \sqrt{-g} R=\frac{M_{pl}^2}{2}\int_{t_1}^{t_2}\left\{\int_{\Sigma_t} N\sqrt{\gamma} ({\mathcal R}+{\mathcal I})dx^3\right\} dt , 
 \end{eqnarray}
 which could be generalized, as a $F$ gravitational theory,  taking the action  
 \begin{eqnarray}
  S_{F;grav}=\frac{M_{pl}^2}{2}\int_{t_1}^{t_2}\left\{\int_{\Sigma_t} N\sqrt{\gamma} F({\mathcal R},{\mathcal I})dx^3\right\} dt. \end{eqnarray}
  
  \
  
  Here, it  is important to stress  that this generalization depends on the chosen foliation. Effectively, for two different slicing, namely $\Sigma_t$ and $\bar{\Sigma}_{\bar{t}}$, the difference between
  ${\mathcal R}+{\mathcal I}$ and $\bar{\mathcal R}+\bar{\mathcal I}$ is a divergence, so, in GR, the action is the same, but when one considers a $F({\mathcal R},{\mathcal I})$ theory the
  action is different from one to other slicing. 
  For this reason, as we have explained in the Introduction, we have to fix a foliation $\Sigma_t$ (the gauge) as in the Cauchy problem, and thus, 
  once the foliation has been chosen, namely $\Sigma_t$, the action is covariant
   for all the coordinate systems where the foliation $\bar \Sigma_{\bar{t}}$ coincides with the chosen one, what always happens for  transformations of the form 
   $\bar{t}=g(t)$, where $g$ is an arbitrary  function independent of the spatial coordinates.

  \
  
  To make physical sense of the theory, the fixed slicing has to be preferred in some sense, { as it could be the frame at rest with respect the
  Cosmic Microwave Background}. To justify our choice,  we first remember the Weyl's postulate, 
  which asserts: "{\it The world-lines of galaxies (on average) form
a bundle of non-intersecting geodesics orthogonal to a series of space-like hyper-surfaces.
This series of hyper-surfaces allows for a common cosmic time and
the space-like hyper-surfaces are the surfaces of simultaneity with respect to
this cosmic time}" \cite{rz1}. So,  the idea behind the  Weyl's postulate is to build up a co-moving reference frame in which
the constituents of the universe are at rest (on average) relative to the co-moving
coordinates. However,  the basis of Weyl postulate (e.g. “non-intersecting (on average) world-lines
of galaxies”) seems questionable if there exists some period in cosmic history when the universe is not filled by dust, for example in the radiation period when the word "galaxies" has to be changed by relativistic gas 
particles, or earlier.

\

To overcome the difficulties of Weyl's postulate and develop it, we follow the Ellis proposal  \cite{ellis1} (see also \cite{rz2}), based on an specific choice of preferred 
fundamental world-lines whose $4$-velocity, which
following the spirit of  the Weyl's postulate has to be intrinsically determined by the constituents of the universe,  is the time-like eigenvector of the stress tensor, i.e., 
satisfyes $T_{\mu}^{\nu}u^{\mu}=\lambda u^{\nu}$. 
For realistic matter, 
it was shown (see page $89-90$ of \cite{haw}) that such family exists and is unique due to the  weak energy condition: $T({\bf v},{\bf v})\geq 0$ for any time-like vector ${\bf v}$.  
What is important is that if this is a $4$-velocity field defined on the whole manifold it implies that its world-lines never cross due to the unicity of the solutions of a differential equation, and thus one has  
a $3+1$
splitting of the spacetime. Finally,
the time could be the corresponding proper time of these world-lines 
defining a preferred synchronous gauge or the 
co-moving slicing defined as the  foliation $\Sigma_t$ 
perpendicular to the world-lines. In fact, as we will show, when the he universe is filled by an scalar field $\phi$, the later  prescription is the better one because in this case the slices 
coincide with the hyper-surfaces $\phi=t=constant$, and thus, the matter component of the universe is not perturbed along a slice.

  \
 
 Note also that, this generalization is completely different from the standard modified $F(R)$ gravity, where the modified Friedmann equation is  a dynamical equation which depends on the second
 derivative of the Hubble parameter (see for instance the reviews \cite{odintsov, odintsov1}), because in our theory, as we will  immediately see, the modified Friedmann equation is a constraint between de 
 Hubble parameter, the energy
 density and the scale factor, that does not contain higher order derivatives.
 
 \
 
As an special case  in cosmology, assuming  the isotropy and homogeneity of the universe,  the  Friedman-Lema{\^\i}tre-Robertson-Walker (FLRW) geometry has to be used, 
whose line element for co-moving observers is given by 
\begin{eqnarray}ds^2=-N(t)dt^2+a^2(t)d\Sigma_t^2.
 \end{eqnarray}

 Now we consider the stress tensor for a perfect fluid $T_{\mu}^{\nu}=(\rho+P)u_{\mu}u^{\nu}+P\delta_{\mu}^{\nu}$ where $P$ is the pressure and ${\bf u}$ is the $4$-velocity of the observers. Since
 $T_{\mu}^{\nu}u^{\mu}=-pu^{\nu}$, this shows that the $4$-velocity is the time-like eigenvector of the stress tensor which leads to the preferred foliation in the FLRW geometry. In the same way, for an scalar
 field the stress tensor is
 $T_{\mu}^{\nu}=\partial_{\mu}\phi\partial^{\nu}\phi-\left(\frac{1}{2}g(\nabla \phi,\nabla \phi)+V(\phi)   \right)\delta_{\mu}^{\nu}$ where $\nabla \phi$ is the gradient of the scalar field. Provided that
 the gradient was time-like, the $4$-velocity ${\bf u}=-\frac{\nabla \phi}{\sqrt{-g(\nabla \phi,\nabla \phi)}}$ is an eigenvector, which for the flat FLRW geometry,  has the form
 ${\bf u}=(\frac{\dot\phi}{N|\dot\phi|},0,0,0)$ and thus leads to the same foliation as the co-moving observers. In fact, when $N=1$,  $t$ is the proper time of the observers and also the slice $t=constant$ are perpendicular
 to the world-lines of the observers, that is, for the FLRW spacetime with $N=1$, the co-moving and synchronous slicing coincide.

 \
 
 Note that, if one takes another foliation $\bar\Sigma_{\bar t}$ with $\bar{t}=g(t,\bf{x})$ and $g$ a function depending of the spatial coordinates $\bf x$,
quantities such as the energy density, the intrinsic or extrinsic curvatures which are constant in every slice, become non-homogeneous in the new slices $\bar{t}=constant$, and thus,  introduce fictitious
 perturbations  when the new slice is a consequence of a simple  coordinate perturbation (see the beginning of section $7$ in \cite{m}).

 \

 Dealing with this preferred slicing, the    extrinsic and intrinsic curvatures are:
 \begin{eqnarray}
 {\mathcal I}=-\frac{6H^2}{N^2}, \qquad {\mathcal R}=\frac{6k}{a^2},
 \end{eqnarray}
 where  $k=0,1,-1$ for a flat, closed and open  slice $\Sigma_t$, respectively.
 
 In that case the  total action would be
 \begin{eqnarray}
 S_{F; tot}= \int_{t_1}^{t_2} NV {\mathcal L}_{grav}dt +\int_{t_1}^{t_2} NV {\mathcal L}_m dt,
 \end{eqnarray}
 where the volume is  $V=a^3$, 
 the gravitational piece of the Lagrangian is ${\mathcal L}_{grav}= \frac{M_{pl}^2}{2}  F({\mathcal R}, {\mathcal I}) $, and the  matter part
 for a barotropic fluid is  ${\mathcal L}_m= -\rho(t)$ \cite{ryan, pazos}, 
 and  ${\mathcal L}_m=  P=\frac{1}{2N^2}\dot{\phi}^2-V(\phi)$
 for a scalar field minimally coupled with gravity, where $P$ is the pressure. As a function of the volume and its derivative one has
 \begin{eqnarray}
 {\mathcal I}=-\frac{2\dot{V}^2}{3N^2V^2}, \qquad {\mathcal R}=\frac{6k}{V^{2/3}}.
 \end{eqnarray}

 Since the action does not depend on the derivative of the lapse function, the  Euler-Lagrange equation becomes
 \begin{eqnarray}
  \partial_N \left(N({\mathcal L}_{grav}+{\mathcal L}_{m})\right)=0,
 \end{eqnarray}
which leads to the well-known Hamiltonian constraint, or equivalently,  for $N=1$ to the modified Friedmann equation 
 \begin{eqnarray}\label{modified}
 -2{\mathcal I} F_{\mathcal I} +  F =\frac{2\rho}{M_{pl}^2}\Longleftrightarrow
 2{H^2}  F_{\mathcal I} +\frac{1}{6}  F =\frac{\rho}{3M_{pl}^2},
  \end{eqnarray}
  where we have denote by $F_{\mathcal I}$ the partial derivative of $F$ with respect  to the extrinsic curvature.
  
 We see that this is a constraint that defines a surface in the space $(a,H,\rho)$, and, in particular,  for the flat FLRW metric a curve in the plane $(H,\rho)$.

 \
 
 The dynamical equation is obtained from the first principle of thermodynamics 
 $d(\rho V)=-PdV$ obtaining the conservation equation
 $\dot{\rho}=-3H(\rho+P)$.  Note also that, when $F({\mathcal R}, {\mathcal I})= {\mathcal R}+ {\mathcal I}$ one obtains the usual Friedmann  equation
 \begin{eqnarray}
 H^2+\frac{k}{a^2}=\frac{\rho}{3M_{pl}^2}.
 \end{eqnarray}

 Finally, dealing with the flat FLRW geometry, since ${\mathcal R}=0$, we will use the notation $f({\mathcal I})\equiv F(0,{\mathcal I})$. Then, given a 
 curve $g({\mathcal I})=\frac{2\rho}{M_{pl}^2}$ in the plane $({\mathcal I}, \rho)$ we can obtain the corresponding $f$ theory solving the first order differential equation 
 (\ref{modified}), where $\rho$ must be replaced by $g({\mathcal I})$. Using the variation of constants method for first order differential equations, the result is given by:
 \begin{eqnarray}\label{x}
 f({\mathcal I})=-\frac{\sqrt{-{\mathcal I}}}{2}\int
  \frac{g({\mathcal I})}{\mathcal I\sqrt{{-\mathcal I}}}d{\mathcal I}.
 \end{eqnarray}

 \subsection{Mimicking holonomy corrected Loop Quantum Cosmology}
 
 Dealing with the flat FLRW geometry with $N=1$, i.e., working, as usual,  in co-moving coordinates, one has   ${\mathcal R}=0$. Then, 
 redefining the Hubble parameter as $\bar{H}=\sqrt{3\rho_c}M_{pl}H$ where $\rho_c$ is a constant with units of energy density, in order that $\bar H$ has the same units as $\rho$,
   in the plane $(\bar{H},\rho)$, the simplest closed curve is a circle,  and since the energy density has to be positive and the Hubble parameter has to be zero when the energy density vanishes, we must choose  a circle
 centered at $(0,\frac{\rho_c}{2})$ with radius $\frac{\rho_c}{2}$, that is,  $\bar{H}^2+\left(\rho-\frac{\rho_c}{2}\right)^2=\frac{\rho_c^2}{4}$, which in the plane $(H,\rho)$ depicts the
 well-known ellipse (see Fig. $1$) that characterizes the holonomy corrected Friedmann equation in   
 Loop Quantum Cosmology (LQC)  (see for instance \cite{singh06,svv06,ccz})
 \begin{eqnarray}
 \frac{{H}^2}{\frac{\rho_c}{12M_{pl}^2}}+\frac{\left(\rho-\frac{\rho_c}{2}\right)^2}{\frac{\rho_c^2}{4}}=1
 \Longleftrightarrow
  H^2=\frac{\rho}{3M_{pl}^2}\left( 1-\frac{\rho}{\rho_c}  \right).
 \end{eqnarray}

\begin{figure}[t!]
    \centering
    \begin{subfigure}[t]{0.49\textwidth}
        \centering
        \includegraphics[width=1\textwidth]{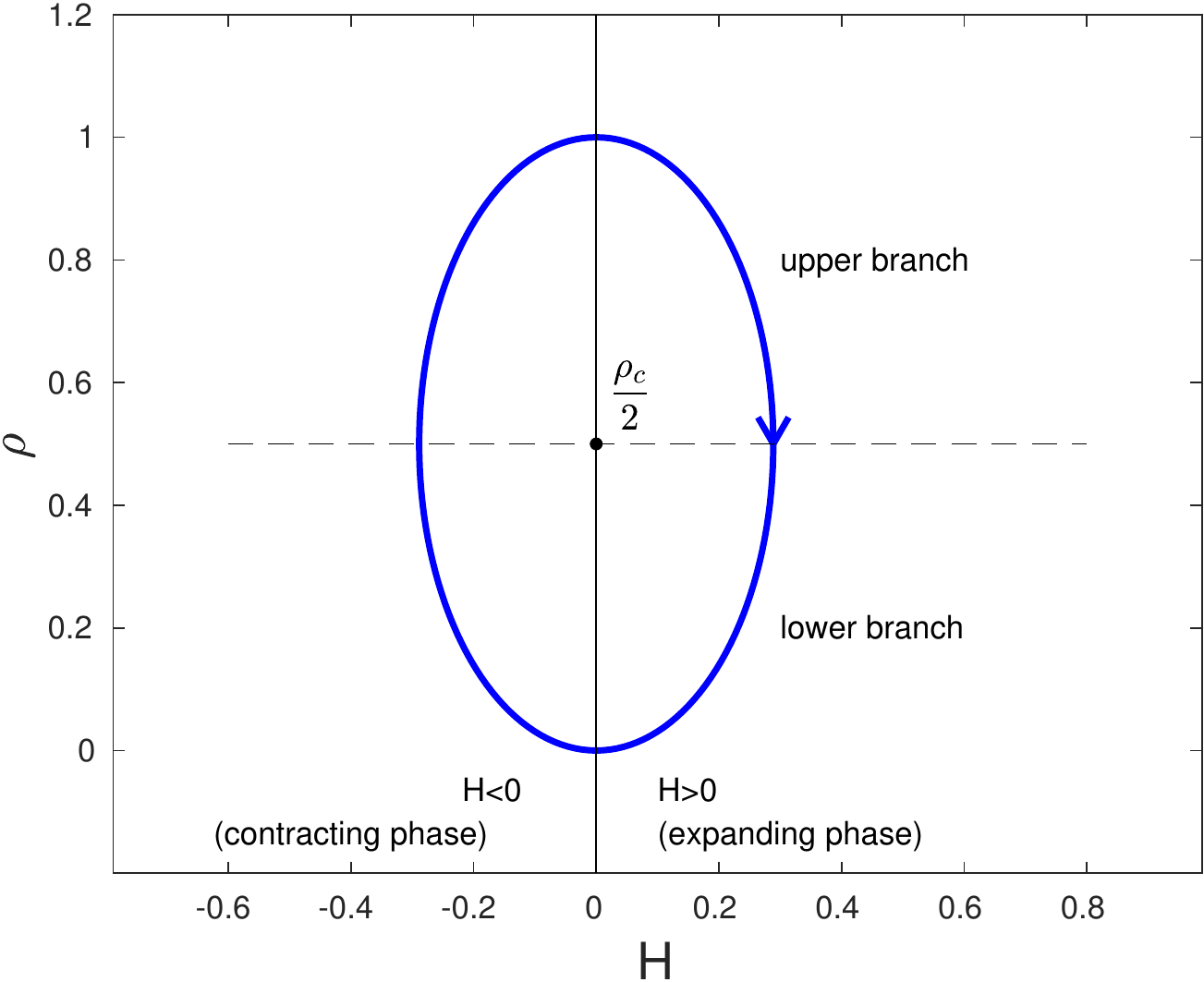}
        \caption{Hamiltonian constraint or equivalently   holonomy corrected Friedmann equation in LQC (units: $\rho_c=M_{Pl}=1$). The arrow shows that the dynamics is ant-clockwise for
        non-phantom matter.}
        \label{f:elipse}
    \end{subfigure}~    
    \begin{subfigure}[t]{0.49\textwidth}
        \centering
        \includegraphics[width=1\textwidth]{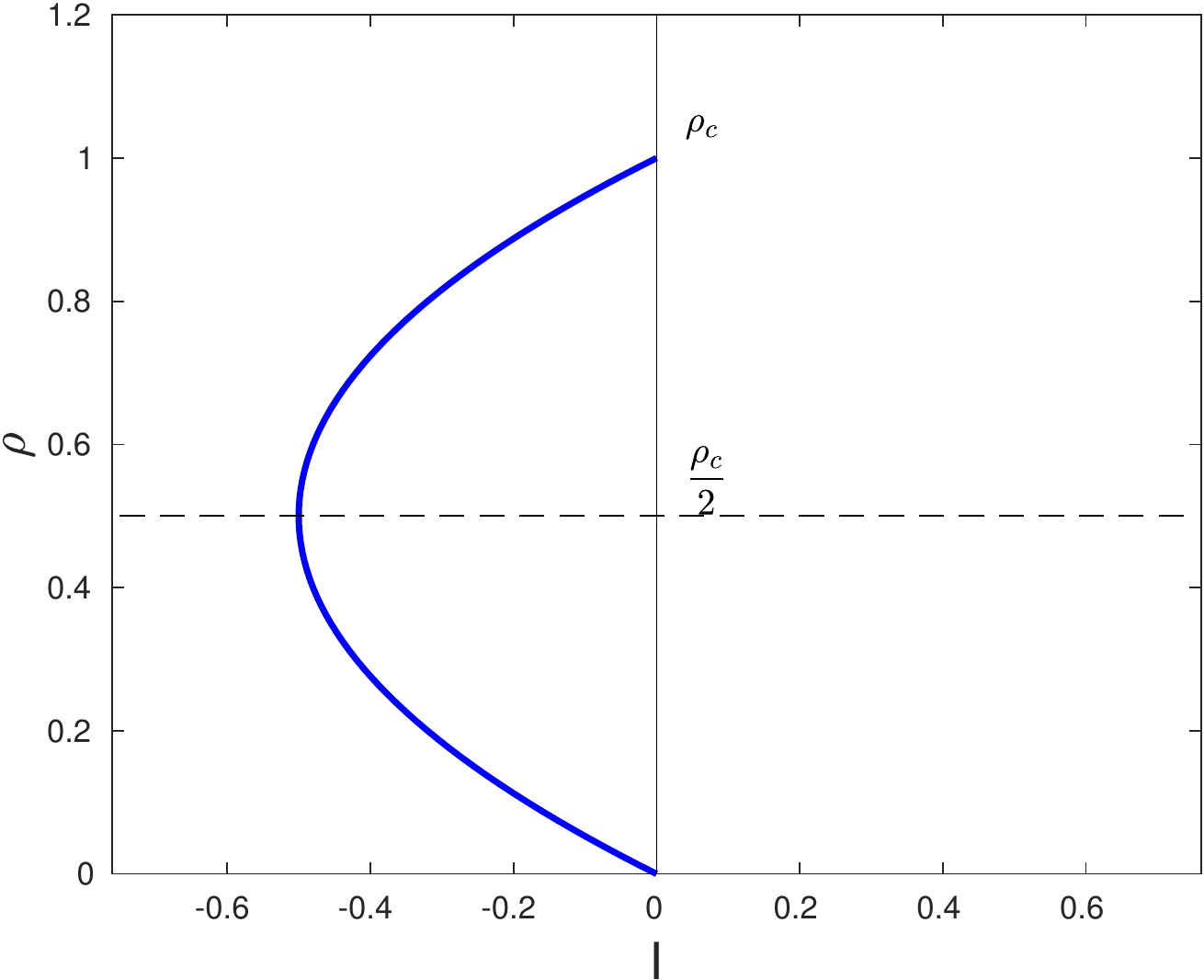}
        \caption{Hamiltonian constraint in the plane $(\mathcal I, \rho)$.}
      \label{f:parabola}
    \end{subfigure}
    \caption{Curve depicting the holonomy corrected Friedmann equation.}
\end{figure}

 The meaning of $\rho_c$, which is called the critical density, is the maximum value of the energy density.
 Since this equation could be written as
 \begin{eqnarray}
 \rho=\frac{\rho_c}{2}\left(1\pm\sqrt{1+\frac{2{\mathcal I}M_{pl}^2}{\rho_c}}\right),
 \end{eqnarray}
 where the sign $-$ correspond to the lower branch of the ellipse, i.e., to values of $\rho$ between $0$ and $\rho_c/2$ and
 the sign $+$ to the upper one, that is, to values between $\rho_c/2$ and $\rho_c$,
 one can see that the energy density is bi-valued  as a function of ${\mathcal I}$, which, {taking into account that mathematically the square root
 is bi-valued},  we will write as
 \begin{eqnarray}
 \rho=\frac{\rho_c}{2}\left(1-\sqrt{1+\frac{2{\mathcal I}M_{pl}^2}{\rho_c}}\right),
 \end{eqnarray}
 where {we use the convention} that the sign positive of the square root corresponds to the lower branch  and the negative to the upper one, i.e.,
 $\rho= \frac{\rho_c}{2}\left(1-\sqrt{1+\frac{2{\mathcal I}M_{pl}^2}{\rho_c}}\right)$ in the lower branch and 
 $\rho= \frac{\rho_c}{2}\left(1+\sqrt{1+\frac{2{\mathcal I}M_{pl}^2}{\rho_c}}\right)$ in the upper one.

 Then, to depict this constraint we will need bi-valued functions $f$,  which could easily be obtained integrating
 the equation (\ref{x}),  \cite{helling, ds09, haro12,ccz}:
 \begin{eqnarray}\label{flqc}
 f({\mathcal I})=\frac{\rho_c}{M_{pl}^2}\left(1- \sqrt{1-s^2}- s\arcsin s  \right),
 \end{eqnarray}
 where
 $s\equiv \sqrt{-\frac{2{\mathcal I} M_{pl}^2}{\rho_c}}=\sqrt{\frac{12H^2 M_{pl}^2}{\rho_c}}$, {belongs in the interval $[0,1]$.}

 Now one has to decide how one choose the values of $f$ in the upper and lower branch. Here, we take two different prescriptions to define the bi-valued function whose picture appears in
 Fig. $2$:
 \begin{enumerate}
 \item  {\bf Prescription 1:}
 We choose  the sign of the square root  positive in the lower branch and negative in the upper one and,
 $\arcsin s\equiv \int_0^s \frac{1}{\sqrt{1-{\bar s}^2}}  {d\bar{s}}$ where the criterium for the sign of the square root is the same.
 \item {\bf Prescription 2:}
  We choose  the sign of the square root  positive in the lower branch and negative in the upper one and, 
$ \arcsin s\equiv \int_0^s \frac{1}{\sqrt{1-{\bar s}^2}}  {d\bar{s}}$ in the lower branch and  $\arcsin s\equiv \int_0^s \frac{1}{\sqrt{1-{\bar s}^2}}  {d\bar{s}}+\pi$
 in the upper one, with the same criterium for the sign of the square root.  
 

 \end{enumerate}

  The advantage of 
the second prescription is that it ensures that the function $f$ is uni-valued when both branches match, i.e. at $s=1$, and in this prescription the function $\arcsin s$ is always positive due to the $\pi$ added in upper branch. Moreover, as we will show dealing with perturbations, the square of the velocity of
sound changes from one prescription to the other.

\

 However, in spite of the authors of \cite{helling, ds09,ccz} arguing that the equation  (\ref{flqc}) gives 
 the $f$ theory that mimics LQC, they only use $f$ as a one-valued function,  meaning that they are only considering the lower branch of the ellipse. We insist in this essential point, to obtain the equivalent background to LQC one
 needs a bi-valued function.
 
\begin{figure}[t!]
        \centering
    \begin{subfigure}[t]{0.49\textwidth}
        \centering
        \includegraphics[width=1\textwidth]{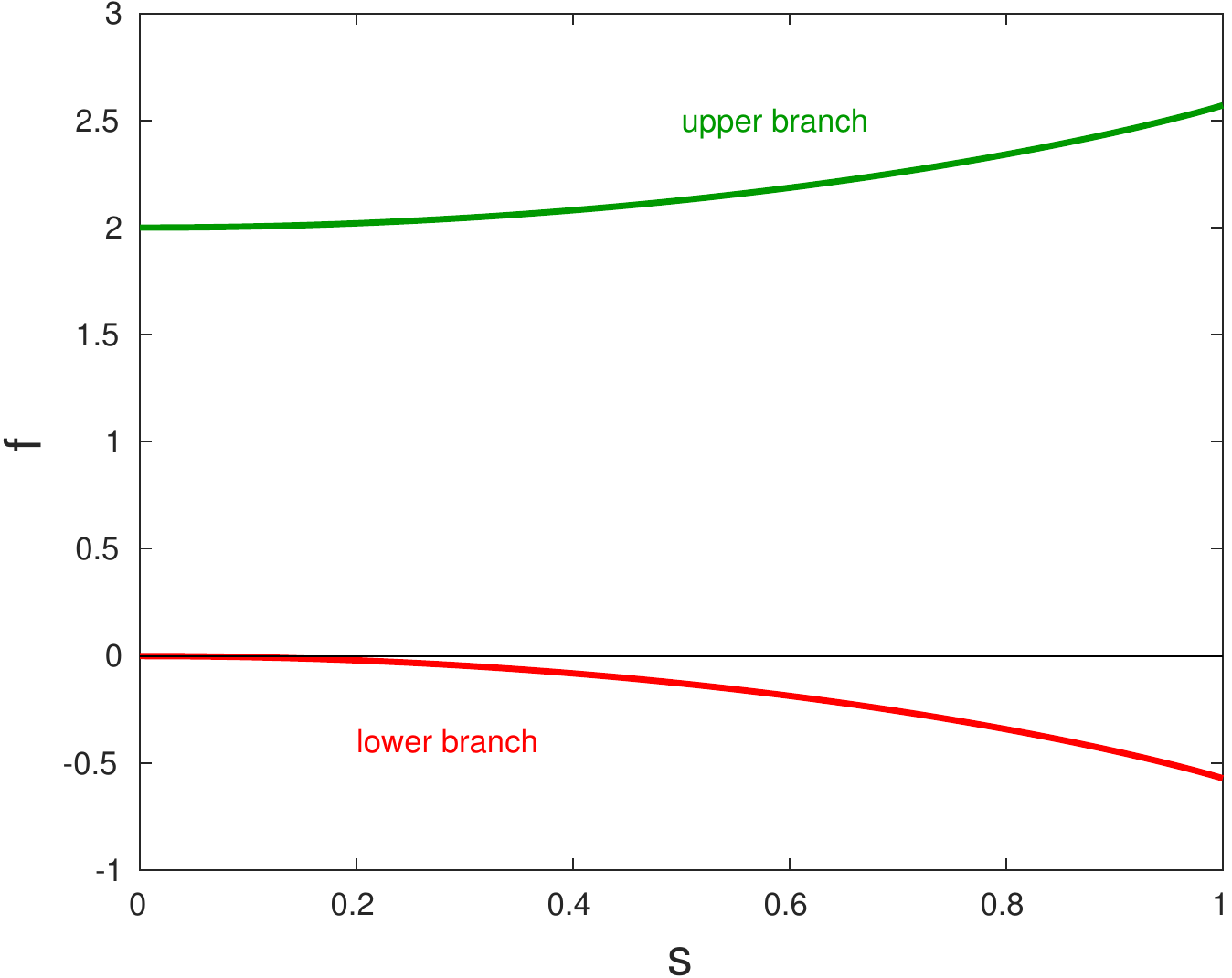}
        \caption{Prescription 1.}
        \label{f:f1}
    \end{subfigure}~
    \begin{subfigure}[t]{0.49\textwidth}
        \centering
        \includegraphics[width=1\textwidth]{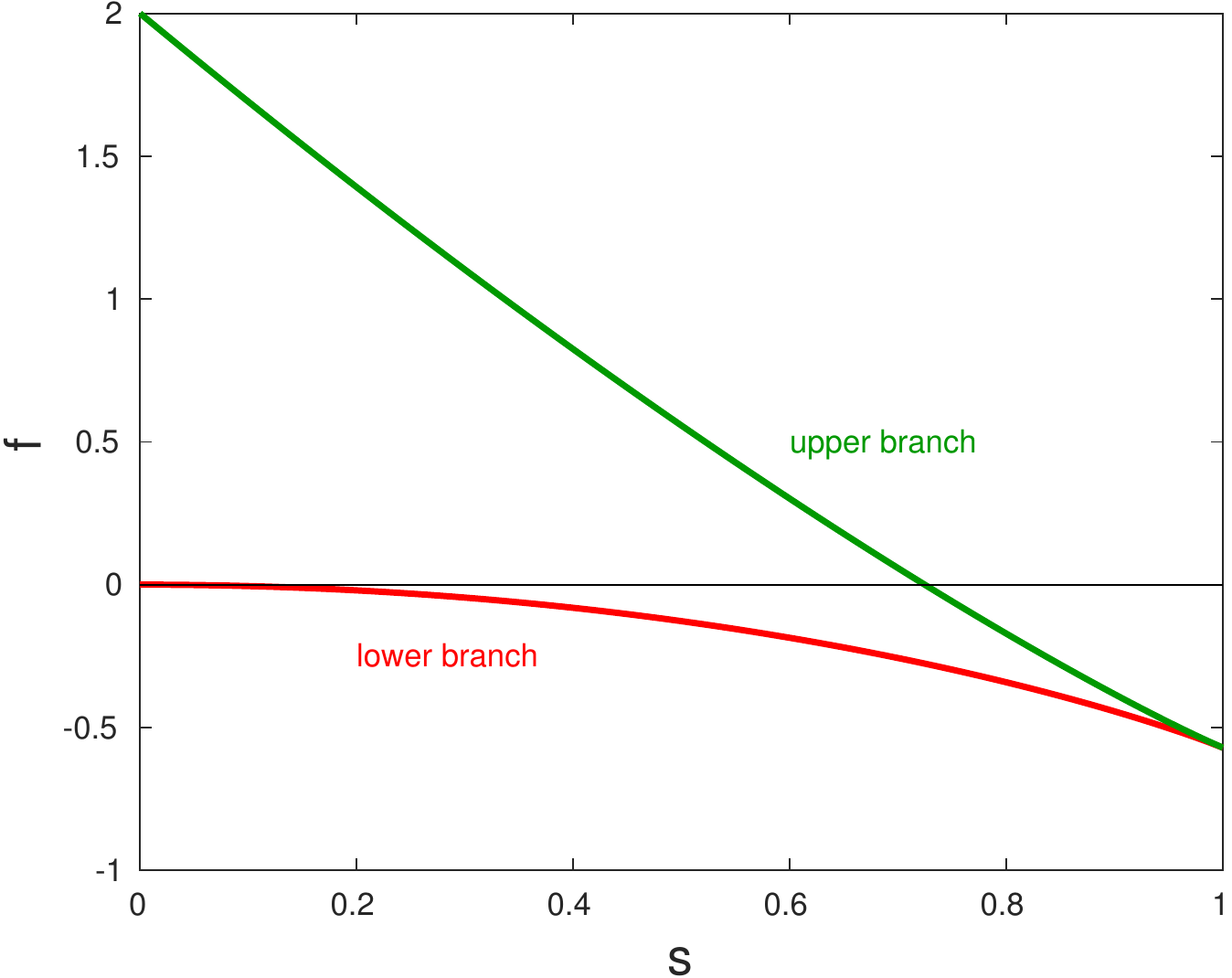}
        \caption{Prescription 2.}
      \label{f:f2}
    \end{subfigure} 
   \caption{Bivalued function $f$ for the upper (green) and lower (red) branches.}
\end{figure}

 \ 
 
 {Some} important final remarks are in order:

 \begin{enumerate}

\item  {As has been pointed out in \cite{mukhanov17}, the bi-valued functions $\sqrt{1-s^2}$ and $\arcsin s$ could be avoided, allowing the variable
$s$ to belongs in the interval $[-1,1]$ (note that by definition it belongs in $[0,1]$) and  performing the transformation
$s=\sin(2\psi)$ with $\psi\in [0,\pi]$. Then, the ellipse is parametrized in the following way
\begin{eqnarray}
H=\sqrt{\frac{\rho_c}{12M_{pl}^2}}\sin(2\psi),\qquad \rho=\rho_c\sin^2 \psi,
\end{eqnarray}
and the total Hamiltoninan
\begin{eqnarray}
{\mathcal H}_{tot}\equiv \dot{V}p_V+\dot{\phi}p_{\phi}-{\mathcal L}_{tot}=-\frac{\rho_c}{M_{pl}^2}(1-\sqrt{1-s^2})+\rho,
\end{eqnarray}
where $p_V=\frac{\partial {\mathcal L}_{tot}}{\partial \dot{ V}}$ and $p_{\phi}=\frac{\partial {\mathcal L}_{tot}}{\partial \dot{ \phi}}$ becomes
\begin{eqnarray}
{\mathcal H}_{tot}=-\frac{\rho_c}{M_{pl}^2}\sin^2\psi+\rho,
\end{eqnarray}
which coincides with the usual Hamiltonian in LQC.

 }
 
 \item 
 At low energy densities a viable theory must coincide with General Relativity, so in the flat FLRW geometry it will be $H^2=\frac{\rho}{3M_{pl}^2}$, 
 meaning that for $H=0$ the energy density vanishes. On the other hand, if one has a bounce,  at the bounce the energy density  does not vanish
when the Hubble parameter is zero. This means that, in a theory with bounces where the modified Friedmann equation is a constraint relating  only $H$ and $\rho$, 
 the energy density as a function of the Hubble parameter has to be a multi-valued function, and thus,   multi-valued functions $f$ are needed to obtain the corresponding modified Friedmann equation.
 
 \item Bouncing cosmologies mimicking LQC could also be obtained, in a  gauge invariant, approach named $F(R,T)$ gravity \cite{harko}, where $T$ denotes the trace of the the stress tensor. In this theory, on the contrary of the one proposed in this manuscript, it is the matter sector which is modified. Dealing with the particular case $F(R,T)=R+ \frac{1}{M_{pl}^2} f(T)$, the modified Friedmann equation, for co-moving observers in the flat FLRW spacetime, becomes \cite{Shabani} 
 \begin{eqnarray}\label{yyy}
 3H^2=\frac{1}{M_{pl}^2}\left[(1+ f'(T))\rho+f'(T){\mathcal L}_{m}-\frac{f(T)}{2}\right].
 \end{eqnarray}
 
 Then, assuming that the universe is filled by an scalar field (this is necessary in order to deal with primordial perturbations) on has ${\mathcal L}_{m}=P$, but in order that the right hand side of  (\ref{yyy}) only depends of $T$, one has to assume that the field
 mimics a perfect fluid with a constant Equation of State parameter $w\equiv \frac{P}{\rho}$. To simplify, we assume that the scalar field mimics  a dust fluid ($w=0 \Longrightarrow T=-\rho$), obtaining the equation
 \begin{eqnarray}\label{yy}
 3H^2=-\frac{1}{M_{pl}^2}\left[(1+ f'(T))T+\frac{f(T)}{2}\right].
 \end{eqnarray}
 
 To obtain  the uni-valued function $f$ leading to the same Friedmann equation as in holonomy corrected LQC, one only has to solve the  first order differential equation
 \begin{eqnarray}
 (1+ f'(T))T+\frac{f(T)}{2} = T\left(1+\frac{T}{\rho_c}\right),
 \end{eqnarray}
 whose solution is  $f(T)=\frac{2}{5\rho_c}T^2$.
 
 Unfortunately,   apart from other difficulties that leads this approach unviable \cite{velter},  the continuity equation differs form the usual one, namely 
 $\dot{\rho}=-3H\rho$  for a pressureless fluid. For $f(T)=\frac{2}{5\rho_c}T^2$, one has
 \begin{eqnarray}
 \dot{\rho}=-\frac{3H\rho}{1-\frac{2\rho}{\rho_c} }\left( 1-\frac{4\rho}{5\rho_c}  \right),
 \end{eqnarray}
 which is singular at $\rho=\rho_c/2$, i.e., when both branches of the ellipse matches. So, only few unnatural an unrealistic bouncing solutions could be obtained (see Section 4 of \cite{Shabani}).
 
 {\item Recently, the function (\ref{flqc}) has been used in the context of mimetic gravity \cite{mukhanov17,langlois17, bss17} 
and also as a function of   the Carminati-MacLenaghan scalars \cite{hp},   
 to reproduce, in a covariant way, the same background as in holonomy corrected
 LQC. A perturbed theory of this approach has not been devoloped yet, although it deserve future investigation.
 
 }


 \end{enumerate}

 \subsection{Gravitational equations in the $F$ theory}

 To get the gravitational  equation we start with 
 the total action 
 \begin{equation}
  S_{F,tot}=\int_{t_1}^{t_2}\left\{\int_{\Sigma_t}N\sqrt{\gamma}\left(\frac{M_{pl}^2}{2}F({\mathcal R},{\mathcal I})+{\mathcal L}_{matt}\right) d^3x\right\} dt,
 \end{equation}
 where we assume the universe is filled by an scalar field  minimally coupled with gravity, whose Lagrangian is
 \begin{eqnarray}
  {\mathcal L}_{matt}= \left(-\frac{\phi_{\mu}\phi^{\mu}}{2}-V(\phi)\right)
  =\left(\frac{\dot{\phi}^2}{2N^2}-\frac{N^i}{N^2}\dot{\phi}\phi_i-\frac{1}{2}\left(\gamma^{ij}-\frac{N^iN^j}{N^2}\right)\phi_i\phi_j-V(\phi)\right).
  \end{eqnarray}

Then,
to obtain the constraints and dynamical equations we have to perform the variation with respect the variables $N,N_i,\gamma_{ij}$ and with respect the scalar field $\phi$.

 \
 
 Taking into account that 
 the components of the extrinsic curvature could be written as follows \cite{Gourgoulhon}
 \begin{eqnarray}
 K_{ij}=\frac{1}{2N}\left(D_iN_j+D_jN_i-\dot{\gamma}_{ij}     \right),
 \end{eqnarray}
 where $D$ is the induced Levi-Civita connection in the space tangent of $\Sigma_t$, that the variation of the 
 intrinsic curvature is
 \begin{eqnarray}
  \delta {\mathcal R}=
 {\mathcal R}_{ij}\delta \gamma^{ij}+D_k\left(\gamma^{ij}\delta \Gamma_{ij}^k-
 \gamma^{kj}\delta\Gamma_{ij}^i  \right),
 \end{eqnarray}
 where ${\mathcal R}_{ij}={\mathcal Ric}(\partial_i,\partial_j)$ are the entries of the Ricci tensor and $\Gamma_{ij}^k$ are the Christoffel symbols, and using the formula 
  \begin{equation}
 \gamma^{km}\left( D_i\delta\gamma_{mj}+ D_j\delta\gamma_{mi} -D_m\delta\gamma_{ij}\right)=2\delta \Gamma_{ij}^k,
 \end{equation}
 and the Gauss divergence theorem 
 one gets (see \cite{tang}, to our knowledge the first instance where these equations were derived):
 \begin{equation}\label{ham}
  F-2{\mathcal I}F_{\mathcal I}=-\frac{2}{M_{pl}^2}\partial_N(N{\mathcal L}_{matt}),
 \end{equation}
\begin{equation}\label{dif}
 D_j(F_{\mathcal I}\pi^{j}_i)=-\frac{N}{M_{pl}^2}\partial_{N^i}{\mathcal L}_{matt},
\end{equation}
\begin{eqnarray}\label{ij}
 -(D^jD_i-\delta^{j}_iD^2)(F_{\mathcal R}N)+
 NG_{F, i}^{j}
 +2NF_{\mathcal I}(K^{mj}K_{im}-Tr (K)K^{j}_i)
 &&\nonumber\\
  -\frac{1}{\sqrt{\gamma}}\gamma_{ki}\partial_t(\sqrt{\gamma}F_{\mathcal I}\pi^{kj})-
 D_k(F_{\mathcal I}(\pi^{k}_iN^j+\pi^{kj}N_i-\pi^{j}_iN^k))
 =\frac{N}{M_{pl}^2}T^{j}_i,
\end{eqnarray}
 where $\pi_i^j$ are the entries of the tensor $\pi\equiv K-Tr(K)\gamma$,
 $D^2=D^kD_k$ is the three-dimensional Laplace-Beltrami operator, $G_F= -\frac{1}{2}F\gamma+F_{\mathcal R}{\mathcal Ric}$ is the three-dimensional modified Einstein tensor,
 and $T_i^j=\frac{2}{\sqrt{\gamma}}\gamma^{mj}\partial_{\gamma_{im}}(\sqrt{\gamma}{\mathcal L}_{matt})$  are the entries of the three-dimensional stress-energy tensor.
 
 \
 
 The equations (\ref{ham}) and (\ref{dif}) are the well-known  Hamiltonian and diffeomorphism constraints, and the equation (\ref{ij}) is the dynamical one.

 \

 For the scalar field, the conservation equation is $-\nabla^2\phi+V_{\phi}=0$, where $\nabla^2=\nabla^i\nabla_i$ is the  Laplace-Beltrami operator in the full manifold ${\mathcal M}$, which in the ADM formalism becomes
 \begin{eqnarray}
  \frac{1}{\sqrt{\gamma}N}\partial_t\left(\frac{\sqrt{\gamma}}{N}\dot{\phi}  \right)-\frac{1}{N}D_i\left(N\phi^i  \right)-
  \frac{1}{N}D_i\left(\frac{N^i}{N}(\dot{\phi}-N^j\phi_j)  \right)+V_{\phi}=0.
 \end{eqnarray}

\

 Finally, once these equations are obtained one can deal with the Cauchy initial value problem as in GR, i.e., given a fixed foliation, in our case the preferred one, one has to choose 
  a slice at some time
 $t_0$, namely $\Sigma$, which is assumed to be a Cauchy hyper-surface), and we have to consider on it the set of variables $(\gamma, K, \phi, \dot{\phi})$ where $\gamma$ is the metric 
 and $K$ the extrinsic
 curvature tensor on $\Sigma$, which has to satisfy the Hamiltonian and Diffeomorphism constraints. Then, to solve the dynamical equations one has to follow, as in GR,  some of the schemes proposed in {\it 
 numerical
 relativity} (see \cite{lenher,isenberg} for reviews).

 \section{Scalar Perturbations}

 {

 In  the previous Section we have shown the lack of covariance of our approach.
  Nevertheless, our theory is gauge invariant at the perturbation level. To prove it, we consider the  general line element
  containing scalar perturbations \cite{riotto}
  \begin{eqnarray}
 ds^2=-(1+2\Psi)dt^2+2a\partial_iB dtdx^i +a^2[(1-2\Phi)\delta_{ij}-2\partial_{ij}^2E-h_{ij}]dx^idx^j,
 \end{eqnarray}  
   where $h$ is a symmetric, traceless and transverse tensor: $h^i_i=\partial_i h^{ij}=0$, and $\Psi$, $\Phi$ and $E$ are scalars \cite{m}.  
   
In order to deal with scalar perturbations, since there are two degrees of freedom we choose the so-called Newtonian or Longitudinal gauge, where the two metric variables $E$ and $B$ vanishes. Then, we obtain  the metric 
 \begin{eqnarray}\label{xxx}
 ds^2=-(1+2\Psi)dt^2+a^2(1-2\Phi)\delta_{ij}dx^idx^j.
 \end{eqnarray}


}


 \
 
 \begin{remark} In General Relativity, i.e., when $F({\mathcal R},{\mathcal I})={\mathcal R}+{\mathcal I}$ and the matter part of the Lagrangian is depicted by an scalar field, one has $\Phi=\Psi$ 
 \cite{m},
 but in general, for a $F$ theory, these functions are different (see  section $7$ of \cite{mfb}). Fortunately, as we will immediately show, if one chooses $F({\mathcal R},{\mathcal I})={\mathcal R}+f({\mathcal I})$ the equality holds.
 Due to this  simplification in the theory,
 we will only consider this kind of $F$ models.
 \end{remark}

 Since in this gauge  one has $N^2=(1+2\Psi)$, $\gamma_{ij}= a^2(1-2\Phi)\delta_{ij}$ and $N^i=0$, the extrinsic curvature is given by
 \begin{eqnarray}
 {\mathcal I}=-\frac{6(H(1-2\Phi)-\dot\Phi)^2}{(1-2\Phi)^2(1+2\Psi)}\approx -6H^2+12H(\Psi H+\dot{\Phi}),
 \end{eqnarray}
 where the symbol $\approx$ means: up to first order.

 \

 On the other hand, since the Christoffel symbols satisfy
 \begin{eqnarray}
 \Gamma^k_{ij}=-\frac{1}{1-2\Phi}
 \left(\delta_{ki}\Phi_j+  \delta_{kj}\Phi_i-\delta_{ij}\delta^{mk}\Phi_m
 \right),
 \end{eqnarray}
 the entries of the  Ricci tensor will be
 \begin{eqnarray}\label{Ricci}
 {\mathcal R}_{ij}=\partial_k\Gamma^k_{ij}-\partial_j\Gamma^k_{ik}+\Gamma^k_{ij}\Gamma^l_{kl}-\Gamma^l_{ik}\Gamma^k_{lj}
 \approx 
 \partial_k\Gamma^k_{ij}-\partial_j\Gamma^k_{ik}=                                                       
 \Phi_{ij}+\delta_{ij}\Delta \Phi,                      
 \end{eqnarray}
 and the scalar curvature will become ${\mathcal R}\approx
 \frac{4}{a^2}\Delta \Phi$,
 where $\Delta=\sum_{i=1}^3 \partial^2_{ii}$ is the standard three-dimensional Laplacian.
 
 \

 Now, in the linear approximation, we will use the following notation:
\begin{eqnarray}
{\mathcal I}={\mathcal I}_0+{\mathcal I}_1,\quad \mbox{with} \quad {\mathcal I}_0=-6H^2, \quad {\mathcal I}_1=12H(\Psi H+\dot{\Phi}),
\end{eqnarray}
\begin{eqnarray}
 \phi=\phi_0(t)+\delta\phi \Longrightarrow V(\phi)=V_0+V_{\phi,0}\delta\phi, \quad \mbox{with}\quad V_0=V(\phi_0),\quad V_{\phi,0}=V_{\phi}(\phi_0).
\end{eqnarray}
\begin{eqnarray}
  g=f-2{\mathcal I}f_{{\mathcal I}}, \quad g_0= g({\mathcal I}_0  )\quad \mbox{ and } \quad g_{{\mathcal I},0}= \partial_{\mathcal I}g({\mathcal I}_0  )\equiv  g_{{\mathcal I}}({\mathcal I}_0  ),
\end{eqnarray}
where the meaning of $g$ is that the equation
$g_0=\frac{2\rho_0}{M_{pl}^2}$, where $\rho_0$ is the non-perturbed energy density, is the
 non-perturbed hamiltonian constraint, i.e., the modified Friedmann equation.

\ 
 
{{} For $i\not= j$,  the equation (\ref{ij}) becomes 
\begin{eqnarray}
-D^jD_i(F_{\mathcal R} N)+NG_{F,i}^j=0,
\end{eqnarray}
 which for our choice $F({\mathcal R}, {\mathcal I})= {\mathcal R}+ f({\mathcal I})$ (essential to obtain the desired result),
 becomes
 \begin{eqnarray}
 -D^jD_i( N)+N{\mathcal R}_i^j=0 \Longrightarrow \partial_{ij}^2(\Phi-\Psi)=0,
 \end{eqnarray}
 which leads to the solution $\Phi=\Psi$.
 Then, the perturbed equations become:}
 \begin{eqnarray}\label{A}
  \frac{1}{a^2}\Delta\Phi+ 3H^2g_{{\mathcal I},0}\Phi+3Hg_{{\mathcal I},0}\dot{\Phi}=
  \frac{1}{2M_{pl}^2}\left(\dot{\phi}_0(\dot{\delta\phi}-\dot{\phi}_0\Phi    )+V_{\phi,0}\delta\phi\right),
 \end{eqnarray} 
  \begin{eqnarray}\label{B}
 - g_{{\mathcal I},0}(H\Phi+\dot{\Phi})  =\frac{1}{2M_{pl}^2}\dot{\phi}_0\delta\phi,
 \end{eqnarray}
\begin{eqnarray}\label{C}
-g_{{\mathcal I},0} \ddot{\Phi}-(4Hg_{{\mathcal I},0}+\dot g_{{\mathcal I},0}     )\dot\Phi-
((3H^2+2\dot{H})g_{{\mathcal I},0}+H\dot g_{{\mathcal I},0} ) \Phi
=\frac{1}{2M_{pl}^2}\left(\dot{\phi}_0(\dot{\delta\phi}-\dot{\phi}_0\Phi    )-V_{\phi,0}\delta\phi\right).
\end{eqnarray}

\

Combining the equations and using the cosmic time one gets
 the dynamical equation for the
 Newtonian potential the following equation, which coincides with the one of GR (see equation (6.48) of \cite{mfb}) when one takes $f({\mathcal I})={\mathcal I}$, 
\begin{eqnarray}\label{holonomy}
 \Phi''-{c}^2_{S}\Delta \Phi+2\left({\mathcal H}- \left(\frac{{{\phi}}_0''}{{{\phi}}_0'}+\epsilon\right)\right)\Phi'
+2\left({\mathcal H}'- {\mathcal H}\left(\frac{{{\phi}}_0''}{{{\phi}}_0'}+\epsilon\right)\right)\Phi=0,
\end{eqnarray}
where the square of the velocity of sound {{} for scalar perturbations} is $c_{S}^2=-\frac{1}{g_{{\mathcal I},0}}$ and we have introduced the notation $\epsilon=-\frac{g'_{{\mathcal I},0}}{2g_{{\mathcal I},0}}$.

\

Now  introducing, as in standard cosmology,
(see for example formulae (8.56)-(8.58) of \cite{m})
the {{} gauge invariant}
Mukhanov-Sasaki (M-S) variables
\begin{eqnarray}\label{19}
 v_S=a\left(\delta\phi+\frac{{\phi}_0'}{{\mathcal H}}\Phi\right), \quad z_S=\frac{a{\phi}_0'}{{\mathcal H}},
\end{eqnarray}
the equation (\ref{holonomy})
becomes, in the Fourier space, the following {{} gauge invariant} M-S equation
\begin{eqnarray}\label{21}
 v_{S,k}'' +\left(c^2_{S}k^2 -\frac{z_S''}{z_S}\right)v_{S,k}=0.
\end{eqnarray}

Some  important final remarks are in order:
\begin{enumerate}


 \item

If one changes ${\mathcal I}$ by ${\mathcal T}$,   where ${\mathcal T}$ is the torsion that appears when one uses the Weitzenb\"ock connection
instead of the Levi-Civita one, these equations are practically the same as the ones that appear in teleparallel $f({\mathcal T})$ gravity \cite{ccdds}. The difference appears in 
equation (\ref{A}), where in $f({\mathcal T})$ gravity the coefficient that multiplies the Laplacian is $\frac{f_{\mathcal T}}{a^2}$, and it is very important because the velocity of sound changes from one 
theory to the other. Effectively, in this theory the equation of the Newtonian potential is the same as  (\ref{holonomy}) but with a velocity of sound given by 
$\tilde{c}_{S,s}^2=-\frac{f_{{\mathcal I},0}}{g_{{\mathcal I},0}}$.  Moreover, in teleparallel $f({\mathcal T})$ gravity the variables $v_S$ and $z_S$ has to be  \cite{haro13,ha14}
\begin{eqnarray}\label{22}
 v_S=\frac{a}{\sqrt{|f_{{\mathcal I},0}|}}\left(\delta\varphi+\frac{{\phi}_0'}{{\mathcal H}}\Phi\right), \quad z_S=\frac{a{\phi}_0'}{\sqrt{|f_{{\mathcal I},0}|}{\mathcal H}},
\end{eqnarray}
in order to recover the M-S equation (\ref{21}), but replacing ${c}_{S}^2$ by $\tilde{c}_{S}^2$.

\item

Applying the theory to the particular  model given in (\ref{flqc}),  we will obtain a theory, which we name {\it Extrinsic curvature LQC},  where
 $-g_{{\mathcal I},0}=\frac{1}{\Omega}$, with $\Omega\equiv 1-\frac{2\rho}{\rho_c}$, and thus,
$c_{S}^2=\Omega$ and $\epsilon= \frac{\Omega'}{2\Omega}$. Then, the equations  (\ref{holonomy}) and (\ref{21}) are identical to the ones obtained in holonomy corrected LQC \cite{cmbg}, {{} which from our point of view supports the consistency of our approach to obtain bouncing cosmologies via modified gravity in the ADM formalism.}

\item On the contrary, in teleparallel LQC, since 
\begin{eqnarray}
 f_{{\mathcal I},0}= \frac{1}{s}\arcsin s,
\end{eqnarray}
where $s= \sqrt{\frac{-2{\mathcal I}M_{pl}^2}{\rho_c}} $,
the square of the velocity of the sound will be
$\tilde{c}_{S}^2=\frac{\Omega}{s}\arcsin s $, and contrary to holonomy corrected and extrinsic curvature LQC,  is always positive if one takes the first prescription below formula (\ref{flqc}),
but taking the second one,  $\tilde{c}_{S}^2$ is positive in the lower branch, negative in the upper one and  unbounded at the bounce. 


\end{enumerate}

\section{Tensor perturbations}

For metric perturbations the line element, in the preferred foliation,  is 
\begin{eqnarray}
 ds^2=-dt^2+a^2(\delta_{ij}-h_{ij})dx^idx^j,
\end{eqnarray}
where, as we have already introduced in the previous section,   $h$ is a symmetric, traceless and transverse tensor: $h^i_i=\partial_i h^{ij}=0$. Due to these properties one has $\sqrt{\gamma}\approx a^3$,
$Tr(K)\approx-3H$ and
${\mathcal I}\cong -6H^2$. Then,  the equation (\ref{ij}) reduces to
\begin{eqnarray}
 2f_{{\mathcal I},0}(K^{mj}K_{mi}+3HK_{i}^j)+G_{F, i}^j-\frac{1}{a^3}\gamma_{ki}\partial_t(a^3f_{{\mathcal I},0}\pi^{kj})=0. 
\end{eqnarray}

\

And  taking into account that ${\mathcal R}_i^j\approx -\frac{1}{a^2}\Delta h_{i}^j$ and ${\mathcal R}\approx 0$,
 the equation for the gravitational waves becomes
\begin{eqnarray}
f_{{\mathcal I},0}\ddot{h}_{i}^{j}+3Hf_{{\mathcal I},0}\dot{h}_{i}^{j}-\frac{1}{a^2}\Delta h_{i}^{j}+\dot{f}_{{\mathcal I},0}\dot{h}_{i}^{j}=0.
\end{eqnarray}

On can see that, this equation differs from the one obtained in $f({\mathcal T})$ gravity, in the coefficient that multiplies the Laplacian.

\

Introducing the variable $v_T=hz_T$ with $z_T={a M_{pl}}\sqrt{|f_{{\mathcal I},0}|}$  and $h$ denoting $h_{i}^{j}$, one obtains, in Fourier space, the M-S equation for tensor perturbations
\begin{eqnarray}\label{tensoreq}
 v''_{T,k}+\left(c_T^2k^2 -\frac{z_T''}{z_T}\right)v_{T,k}=0,
\end{eqnarray}
where now the square of the velocity of the sound is equal to $c^2_T=\frac{1}{f_{{\mathcal I},0}}$.

\

The variables $v$, $z_T$ are the same used en teleparallel $f({\mathcal T})$ gravity, and the M-S for gravitational waves only differs in the square of the velocity of sound, which in teleparallelism is  always $1$. 

\




Now,
if we consider the function that leads to the same background as holonomy corrected LQC, one has
$ f_{{\mathcal I},0}= \frac{1}{s}\arcsin s,$
what implies that, if one takes the first prescription below (\ref{flqc}) the square of the velocity of the sound $c^2_T=\frac{1}{f_{{\mathcal I},0}}$
is positive in the lower branch of the ellipse and negative and the upper one, as in holonomy corrected LQC, although it is different because in LQC for both
kind of perturbations, scalar and tensors, one has $c^2_T=\Omega$. Moreover, in  extrinsic curvature LQC following this prescription, the square of the velocity  is discontinuous 
when the branches match. This does not happen if one chooses the second prescription where  $c^2_T$ is always positive and continuous.

However, 
the main difference  with holonomy corrected LQC appears in the definition of $z_T$, because in the former case  one has \cite{gbg}
$z_T=\frac{a M_{pl}}{\sqrt{c_T^2}}=\frac{a M_{pl}}{\sqrt{\Omega}}$, which becomes complex in the upper branch, {{}  being  an exotic feature of holonomy corrected LQC, and reducing the ratio of
tensor to scalar perturbations to negligible values in the matter bounce scenario (see next section) \cite{wilson, ha14}. This does not happen in theories such as GR, $F(R)$ or  $F({\mathcal T})$ gravity, and neither does it in our approach defining $z_T=\frac{a M_{pl}}{\sqrt{|c_T^2|}}={a M_{pl}}\sqrt{\frac{|\arcsin s|}{s}}$. But, if one chooses
 $z_T=\frac{a M_{pl}}{\sqrt{c_T^2}}={a M_{pl}}\sqrt{\frac{\arcsin s}{s}}$ (which could be done because the equation (\ref{tensoreq}) is the same for both definitions of $z_T$), for the first
 prescription of $f$, since the square of the velocity of sound becomes negative in the upper branch,  $z_T$ will be complex in the upper branch as in holonomy corrected LQC.
 This will reduce the tensor/scalar ratio to negligible values in the case of the matter bounce scenario, which does not happen with the second prescription, because in that case $c_T^2=\frac{s}{\arcsin s}$ is always positive.

}

\

{{} Summing up, the M-S for scalar and tensor perturbation in {\it Extrinsic curvature LQC} are:

\begin{eqnarray}
 v''_{S,k}+\left(c_S^2k^2 -\frac{z_S''}{z_S}\right)v_{S,k}=0, \qquad v''_{T,k}+\left(c_T^2k^2 -\frac{z_T''}{z_T}\right)v_{T,k}=0,
\end{eqnarray}
where, if one considers the model that leads to the same background as holonomy corrected LQC, i.e., given by the function (\ref{flqc}) and we use the second prescription,
the square of the velocity of sound is $c_S^2=\Omega=1-\frac{2\rho}{\rho_c}$ for scalar perturbations and $c_T^2=\frac{s}{\arcsin s}$ for the tensor ones. Then, since in holonomy corrected LQC one has $c_S^2=c_T^2=\Omega$ we see that both approaches coincide for scalar perturbations, but for tensor ones they differ in the square of the velocity
of sound, where in holonomy corrected LQC the sign of the square of the velocity of sound changes its sign passing from one branch of the ellipse to the other one, but in Extrinsic curvature LQC it is always positive and in fact, for tensor perturbations,  as we already explained, our approach is closer to  $F({\mathcal T})$ gravity. In conclusion,
Extrinsic curvature LQC leads to the same scalar perturbation equations that holonomy corrected LQC, and very close equations to  $F({\mathcal T})$ gravity for tensor
perturbations.

}

\section{The matter bounce scenario in theories with the same background as LQC}

In this scenario the universe is filled by dust matter, whose background in LQC,  can be mimicked by a scalar field whose potential  is \cite{miel}
\begin{eqnarray}
V(\phi_0)=2\rho_c\frac{ e^{-\frac{\sqrt{3}\phi_0}{M_{pl}}}   }
{\left(1+  e^{-\frac{\sqrt{3}\phi_0}{M_{pl}}}    \right)^2},
\end{eqnarray}
where $\phi_0$ is the homogeneous part of the scalar field, i.e., the unperturbed part. The conservation equation is given by
\begin{eqnarray}\label{conservation}
\ddot{\phi}_0+3H_{\pm}\dot{\phi}_0+V_{\phi_0}=0,
\end{eqnarray}
where $H_{\pm}=\pm\sqrt{\frac{\rho_0}{3M_{pl}^2}\left( 1-\frac{\rho_0}{\rho_c}   \right)}$ are the values of the Hubble parameter in the contracting and expanding phase, and $\rho_0$ is the unperturbed
energy density
$\rho_0=\frac{\dot{\phi}_0^2}{2}+V(\phi_0)$. Thus,
for each  of the infinite solutions of (\ref{conservation}) one obtains a different background $H(t)$.  On the other hand, there is an analytic solution given by
\begin{eqnarray}\label{an}
\phi_0(t)=\frac{2M_{pl}}{\sqrt{3}}\ln \left(   \sqrt{\frac{3\rho_ct^2}{4M_{pl}^2}}+  \sqrt{\frac{3\rho_ct^2}{4M_{pl}^2}+1} \right),
\end{eqnarray}
which leads to the following analytic background 
\begin{eqnarray}
H(t)=\frac{\frac{\rho_c t}{2}}{\frac{3\rho_ct^2}{4}+M_{pl}^2}.
\end{eqnarray}

\begin{figure}[t!]
    \centering
    \begin{subfigure}[t]{1\textwidth}
        \centering
        \includegraphics[width=0.6\textwidth]{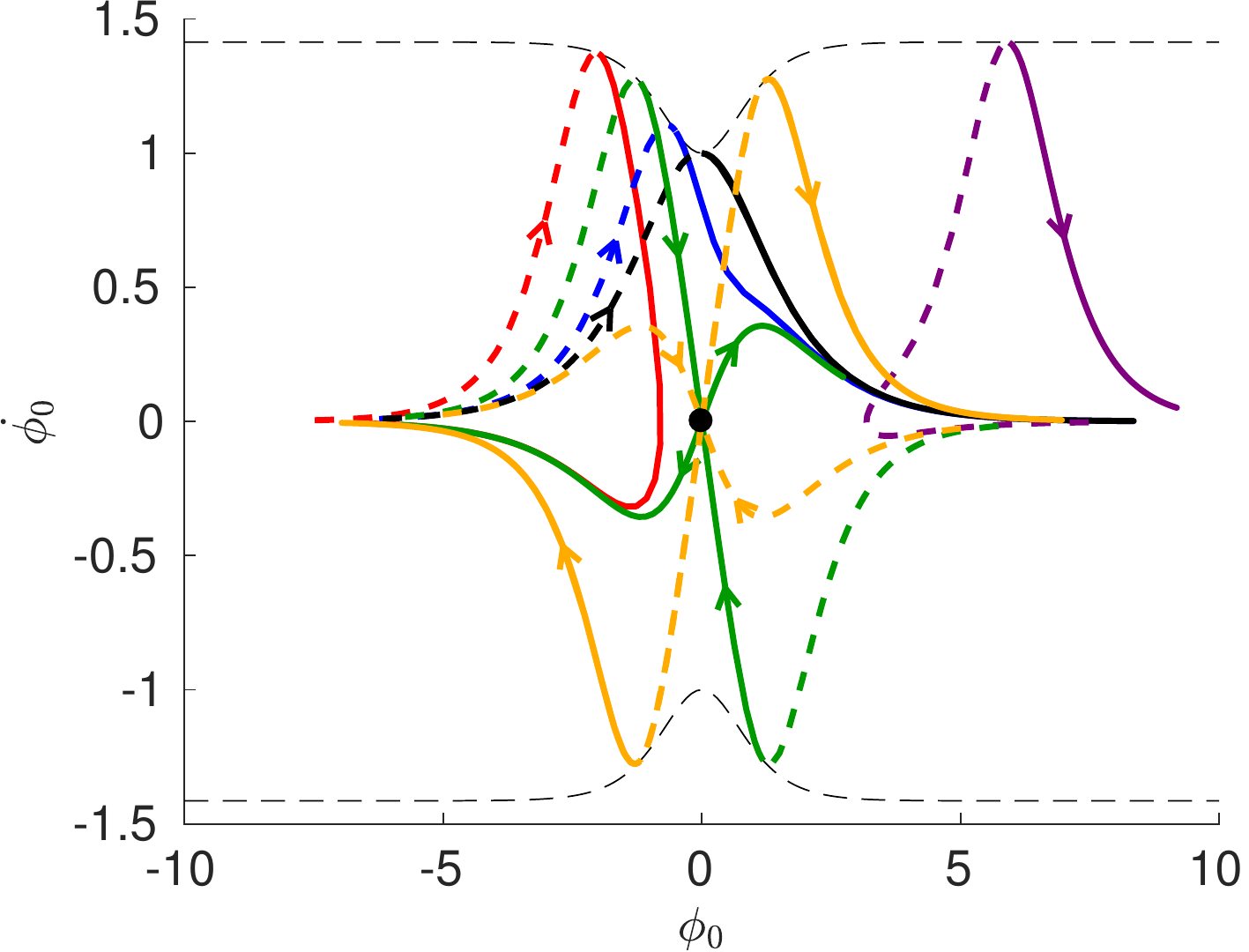}
        \caption{Phase portrait of the system, with curves bouncing from 
$H<0$ to $H>0$ at the upper bounce curve where $\dot{\phi_0}>0$: the unstable and stable 
varieties for $H>0$ (green), and $H<0$ (orange) of the critical point at (0,0); orbits bouncing to the left (red) and to the right (blue) of the
stable variety; the analytic orbit (black); an orbit bouncing past the unstable variety (purple). Orbits with $H>0$ are plotted as continuous 
lines; those with $H<0$ as discontinuous lines.}
        \label{f:retratfase}
    \end{subfigure}
    
    \begin{subfigure}[t]{1\textwidth}
        \centering
        \includegraphics[width=0.5\textwidth]{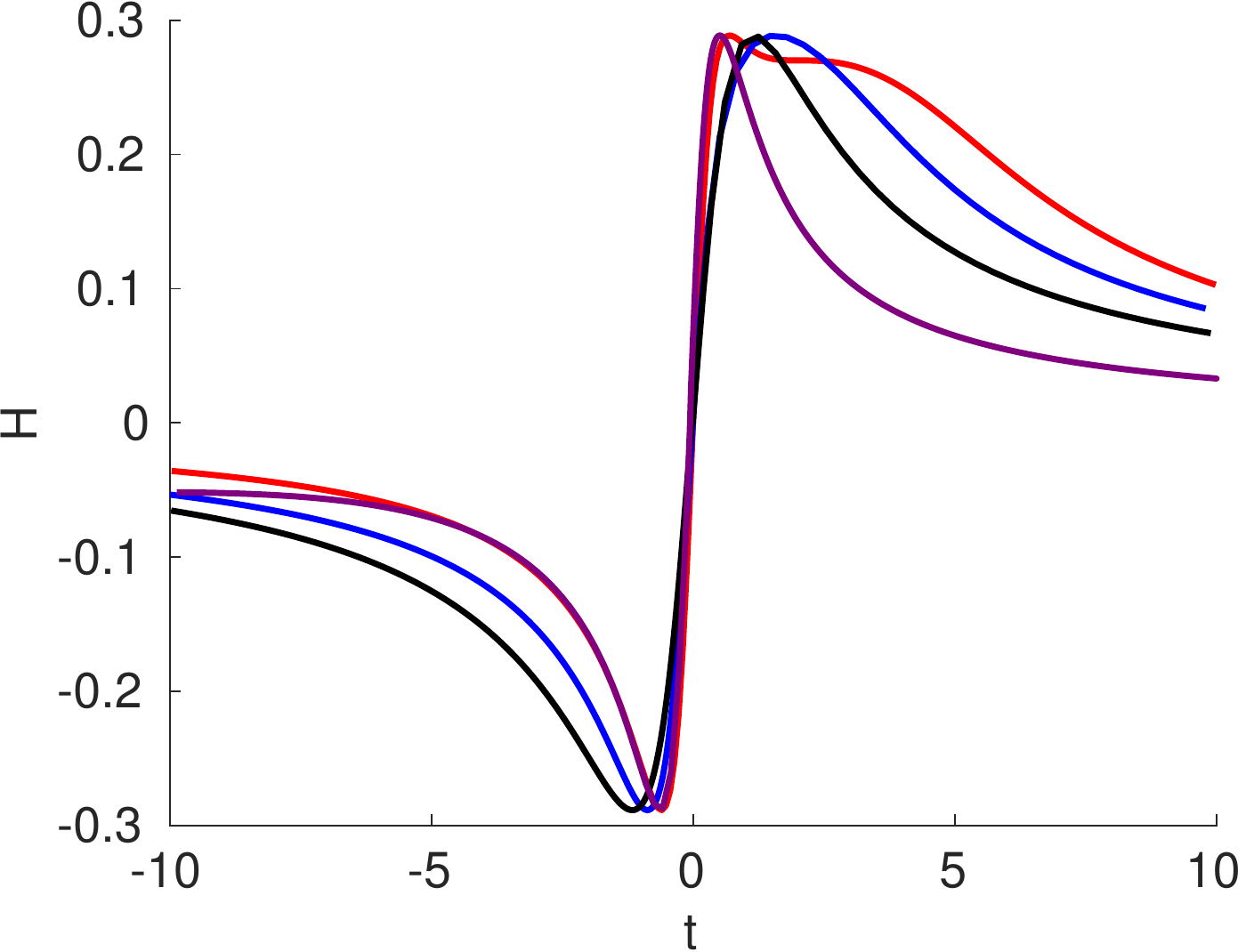}
        \caption{$H(t)$ for the orbits in Fig. \ref{f:retratfase} 
bouncing left of the stable variety (red), between the stable and unstable varieties (blue), right of the unstable variety (purple), and for the analytic orbit (black).}
      \label{f:grafsH}
    \end{subfigure}
    \caption{The extrinsic curvature LQC model: possible backgrounds 
(black: analytic solution; red, blue, purple: qualitatively different
alternatives).}
\end{figure}

The other solutions have to be calculated numerically, and lead to different backgrounds (see Figures \ref{f:retratfase}, \ref{f:grafsH},
also the figures  $8$ and $9$ of \cite{haa} and the corresponding explanation).

Dealing with the power spectrum,  for scalar perturbations it is given by the formula \cite{riotto}
\begin{eqnarray}
 P_S(k)=\frac{k^3}{2\pi^2}|\zeta_k|^2,
\end{eqnarray}
where, as we have already explained in Section III,  $\zeta_k\equiv \frac{v_k}{z_S}$ is the curvature fluctuation  in co-moving coordinates. In the case of the matter bounce scenario, taking at early 
times the scale factor $a(t)= \left( \frac{3\rho_ct^2}{4M_{pl}^2}  \right)^{1/3} $,  one has \cite{ha14a}
\begin{eqnarray}
 P_S(k)=\frac{3\rho_c^2}{\rho_{pl}}\left(\int_{-\infty}^{\infty}\frac{1}{a(t)z^2_S(t)}dt\right)^2,
\end{eqnarray}
and for tensor perturbations
\begin{eqnarray}
 P_T(k)=\frac{\rho_c^2}{\rho_{pl}}\left(\int_{-\infty}^{\infty}\frac{1}{a(t)z^2_T(t)}dt\right)^2,
\end{eqnarray}
 where $\rho_{pl}\equiv 64\pi^2 M_{pl}^4$ is Planck's energy density.

Therefore, the ratio of tensor to scalar perturbations is
\begin{eqnarray}
r=\frac{\left(\int_{-\infty}^{\infty}\frac{1}{a(t)z^2_T(t)}dt\right)^2}{3\left(\int_{-\infty}^{\infty}\frac{1}{a(t)z^2_S(t)}dt\right)^2}.
\end{eqnarray}

\

Using the analytic solution (\ref{an}), and the previous notation $\Omega=1-\frac{2\rho}{\rho_c}$ and $s=\sqrt{\frac{12H^2 M_{pl}^2}{\rho_c}}$, one obtains for the first prescription of $f$:
\begin{enumerate}
 \item Holonomy corrected LQC:
 In this case one has $z_S=a\frac{\dot\phi_0}{H}$ and $z_T=\frac{a M_{pl}}{\sqrt{\Omega}}$, which leads to
 $P_S(k)=\frac{\pi^2\rho_c}{9\rho_{pl}}$ (see \cite{wilson}) and $P_T(k)=0$. Thus,  $r=0$.
 \item Teleparallel LQC:
  In this case one has $z_S=a\frac{\dot\phi_0}{H}\sqrt{\frac{s}{|\arcsin s|}}$
  and $z_T={a M_{pl}}\sqrt{\frac{|\arcsin s|}{s}}$, which leads to
 $P_S(k)=\frac{16 C^2\rho_c}{9\rho_{pl}}$, where $C=\sum_{n=1}^{\infty}\frac{(-1)^n}{(2n+1)^2}\cong 0.91596... $ is Catalan's constant, and 
 $P_T(k)=\frac{16}{3}Si^2(\frac{\pi}{2})\frac{\rho_c}{\rho_{pl}}$ where $Si(x)=\int_0^x\frac{\sin z}{z}dz$ is the Sine integral function, and thus, $r=3\left(\frac{Si(\frac{\pi}{2})}{C}\right)^2\cong
 6.7$.
 \item Extrinsic curvature LQC: 
 In this case one has $z_S=a\frac{\dot\phi_0}{H}$ and $z_T={a M_{pl}}\sqrt{\frac{|\arcsin s|}{s}}$,
 which leads to
 $P_S(k)= \frac{\pi^2\rho_c}{9\rho_{pl}}  $ and $P_T(k)=\frac{16}{3}Si^2(\frac{\pi}{2})\frac{\rho_c}{\rho_{pl}}$, and thus,
 $r=\frac{48}{\pi^2}Si^2(\frac{\pi}{2})\cong 9.1$.
\end{enumerate}

\ 

\begin{figure}[t!]
    \centering
    \begin{subfigure}[t]{0.49\textwidth}
        \centering
        \includegraphics[width=1\textwidth]{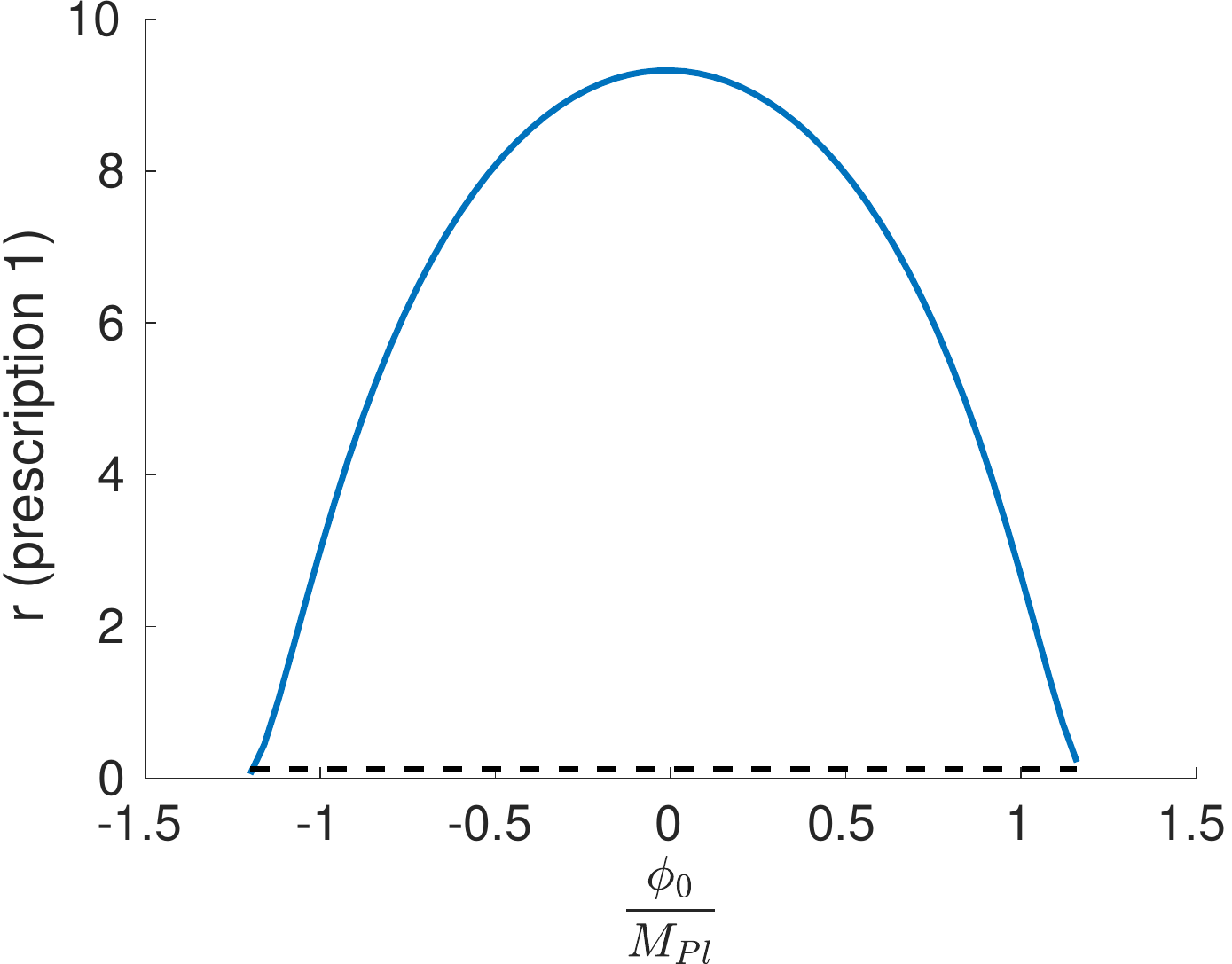}
        \caption{For prescription 1 of $f$.}
        \label{f:r_presc1}
    \end{subfigure}~    
    \begin{subfigure}[t]{0.49\textwidth}
        \centering
        \includegraphics[width=1\textwidth]{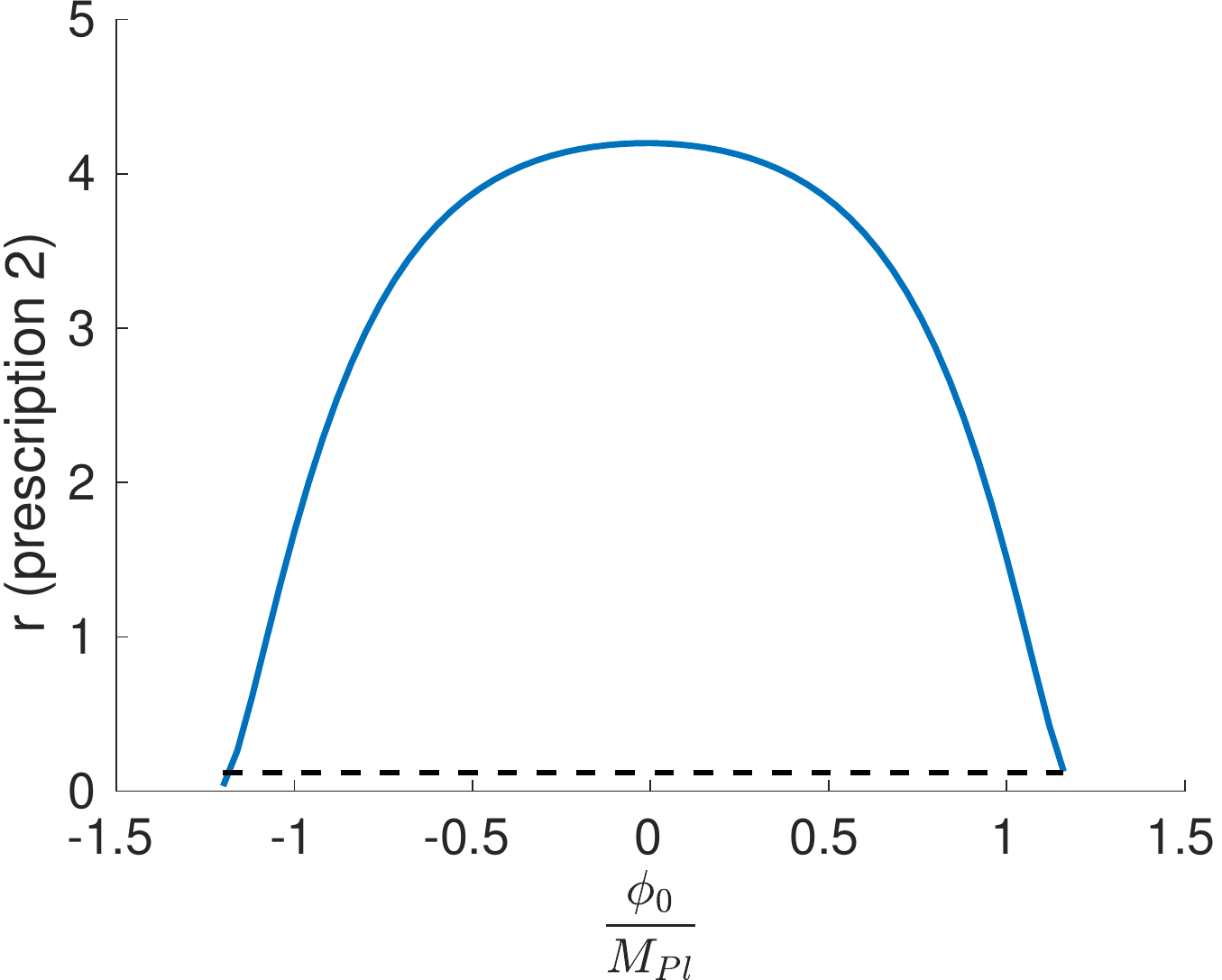}
        \caption{For prescription 2 of $f$.}
      \label{f:r_presc2}
    \end{subfigure}
    \caption{Tensor/scalar ratio $r$ for backgrounds solution 
to Extrinsic curvature LQC Eq. (\ref{conservation}). 
The backgrounds are indexed according to the value of $\phi_0$ at bounce. The Planck-BICEP2 bound $r \leq 0.12$ is indicated.}
\label{f:r}
\end{figure}

We can see that in the case of Teleparallell and Extrinsic curvature LQC, the analytic solution (\ref{an}) leads to a very high value of the tensor/scalar ratio. Fortunately, as already happens for the case of Teleparallel LQC \cite{ha14}, some backgrounds in Extrinsic curvature LQC lead to a ratio of tensor to scalar perturbations that satisfy the constraint $r\leq 0.12$
provided by 
the joint analysis of BICEP2/Keck Array and Planck teams \cite{planck}.
Indeed, the authors' numerical computations, illustrated in Figure \ref{f:r},
show that for backgrounds bouncing at $\frac{\phi_0}{M_{pl}} \in [-1.22,-1.20] \cup 
[1.16, 1.20]$ the tensor/scalar ratio $r$ is below 0.12, while the scalar power spectrum $P_S(k)$ is still finite, requiring a fine-tuning of the value of the scalar field at the bounce point.
The values of $P_S(k)$ in the admissible domain $[-1.22,-1.20] \cup 
[1.16, 1.20]$ grow to infinity abruptly at the domain's upper and lower bounds $\frac{\phi_0}{M_{pl}}=-1.22,1.20$, and descend to $16\frac{\rho_c}{\rho_{pl}}$  at the inner bounds 
$\frac{\phi_0}{M_{pl}}=-1.20,1.16$. Consequently, since in Extrinsic curvature LQC, the critical energy density is a parameter of the theory, in order to match  with  the observational data $P_S\cong 2\times 10^{-9}$  \cite{wmap} we have to choose 
$\rho_c\sim 10^{-10}\rho_{pl}$ and  fine tune the  value of  the backgrounds  at the bouncing time in order to  satisfy that  
$\frac{\phi_0}{M_{pl}}$  (at bounce) was  extremely  close to $-1.20$ or $1.16$.  However in holonomy corrected LQC
the critical  energy density has a determined value 
obtained relating the black hole
entropy in LQC with the Bekenstein-Hawking entropy formula \cite{meissner},  which is approximately $\rho_c\cong 0.4 \rho_{pl}$     disagreeing  with   the value obtained in the matter bounce scenario.

\ 

Finally, note that using the second prescription  (see below (\ref{flqc})) to define  the bi-valued  function $f$, scalar perturbations in teleparallel LQC 
are not well defined because the power spectrum of scalar perturbations diverges, this is the reason why the first prescription is used in \cite{ha14,ha14a}. On the contrary, in extrinsic curvature LQC both kinds of perturbations are well defined:
the scalar ones are independent of the prescription and for tensor ones, in both prescriptions, the power spectrum is of the same order. In fact, for the analytic solution
(\ref{an}) we have obtained $P_T(k)=\frac{4}{3}Si^2({\pi})\frac{\rho_c}{\rho_{pl}}$, leading to a tensor/scalar ratio equal to $r=\frac{12}{\pi^2}Si^2({\pi})\cong 4.1$. Figure \ref{f:r} illustrates the comparison of the 
values of the tensor/scalar ratio $r$ for both prescriptions of $f$;
the backgrounds for which $r \leq 0.12$ turn out to be almost exactly the
same.

\subsection{Improvement of the matter bounce scenario: the matter-ekpyrotic bounce scenario}

As we have seen above when calculating the power spectrum,  dealing with the LQC background, it is of the order $\frac{\rho_c}{\rho_{pl}}$, then to match with the observational value $ P_S(k)\sim 2\times 10^{-9}$, one has
to take $\rho_c\sim 10^{-10} \rho_{pl}$.  This is not a problem  neither for teleparallel LQC  nor for our extrinsic curvature LQC because in both theories the critical density is only a parameter, however this  is in contradiction 
 with  the value of the 
the critical energy density  in holonomy corrected LQC which has  the  determined value   $\rho_c\cong 0.4 \rho_{pl}$.

\

In the same way, 
the matter bounce scenario leads to a flat power spectrum, i.e., the spectral index of scalar perturbations is exactly equal to $1$,  which does not agree with the recent observational results provided by 
the joint analysis of BICEP2/Keck Array and Planck teams \cite{planck}. This problem
could easily overpassed assuming that the universe is not exactly matter dominated, rather with an Equation of State $P=-\epsilon \rho$ with $\epsilon\approx 0.003$ \cite{wilson}. However, the problem is not solved at all because,
as we have already shown, in order to have a  tensor/scalar ratio satisfying the constraint $r\leq 0.12$ we have to fine tune the value of the scalar field at the bounce.
 For this reason, a more sophisticated improvement seems necessary. One way comes from the so-called matter-ekpyrotic bounce scenario introduced in \cite{ceb},  developed in \cite{glm,cbp}, and studied for the particular of the LQC background in \cite{cw,haa}. In this new scenario, the universe evolves at very early times in the contracting phase from a matter domination regime to an ekpyrotic one, which leads, in the case of the LQC background with  the model proposed in \cite{cw},  to a power spectrum of scalar perturbations of the order $H_E^2/M_{pl}^2$, where $H_E$ is the value of the Hubble parameter at the phase transition from matter domination 
to ekpyrotic phase. Moreover, in the same framework of LQC, the improved model studied in \cite{haa} leads to the same power spectrum of scalar perturbations as in \cite{cw}, with a 
spectral index and running entering in its observational
$1$-dimensional marginalized domain at $2\sigma$ C.L..

\

On the other hand,  apart from solving the 
well-known
Belinsky-Khalatnikov-Lifshitz (BKL) instability: the efective energy density of primordial
anisotropy scales as $a^{-6}$ in the contracting phase \cite{bkl}, the  more important thing is that the phase transition provides the mechanism to create enough particles, via gravitational particle production, to reheat and thermalize 
the universe in order to  math with the hot
Friedmann universe \cite{qcb, he15}.

  \section{Conclusions}

 We have presented a theory based in the ADM formalism where, given a pre-determined slicing, the Lagrangian of the gravitation sector is proportional to  the intrinsic curvature
 plus a function $f$ depending on the extrinsic  one. The a priori 
slicing is the price one has to pay in order to generalize, using the ADM formalism,
 GR to an $f$-theory, and similarly 
 in teleparallelism where a fixed orthonormal basis in the tangent bundle has to be  chosen,
 or in holonomy corrected LQC where the  Asthekar connection is replaced by a convenient sinus function.
 
 \

In the work we have tried to justify the choice of a preferred foliation based on an improvement of Weyl's postulate, 
 where the $4$-velocity of the preferred observers
 is the time-like eigenvector of the stress tensor, which goes precisely in the spirit of Weyl's postulate, where the matter content of the universe must fix a preferred frame. Fortunately, we have shown that
the equations dealing
 with perturbations are gauge invariant, as in General Relativity, teleparallelism or holohomy corrected LQC.

 \
 
 At the level of the background, i.e., working in the flat FLRW spacetime with the usual co-moving frame, we show that in order to have a bounce, the use of multivalued functions $f$  is mandatory , and the
 simplest one is the one that leads to the holonomy corrected Friedmann equation in LQC. For a general metric, using the variational principle we have obtained, in the case of  a general $f$-theory, the 
 Hamiltonian and 
 diffeomorphism constraints and the dynamical equations, which have been used to calculate the equations of perturbations in this theory.

 \
 
 We have compared them with the ones obtained in teleparallel $f({\mathcal T})$ gravity,
 being {$\mathcal T$}  the torsion provided by the Weitzenb\"ok connection, 
 showing that for tensor perturbations the M-S equation coincides, and for scalar perturbations it only differs in the square of the velocity of sound. Moreover,
  for the particular case of the $f$ which leads to the same background of LQC, we have compared them with the perturbed equations
 of the holonomy corrected LQC, obtaining that for scalar perturbations, the M-S equation is the same, but for tensor perturbations it is 
 completely different. 
 
 \
 
 Finally, we have applied  the above results  to the matter bounce scenario in LQC, calculating the power spectrum for scalar and tensor perturbations, and showing numerically that, by choosing carefully 
  some backgrounds, 
  the ratio of tensor to scalar perturbations, namely $r$,  satisfies the observational  
 constraint $r\leq 0.12$ provided by 
 the joint analysis of BICEP2/Keck Array and Planck teams \cite{planck}.

\section*{Acknowledgments} 
This investigation has been supported in part by MINECO (Spain), projects  MTM2017-84214-C2-1-P
 and MTM2015-69135-P, and by the Catalan Government 
projects 2017-SGR-247 and 2014-SGR-634.
Finally, the authors are very grateful
the reviewer for his constructive suggestions that result in an improved version of the manuscript 
both in quality as well as in presentation.

\end{document}